\documentclass[12pt]{article}
\usepackage{amsmath}
\usepackage{epsfig}
\usepackage{amssymb}
\usepackage{hyperref}
\usepackage{bbold}
\usepackage{braket}
\input{epsf}
\def\be{\begin{equation}}
\def\ee{\end{equation}}
\def\beq{\begin{equation}}
\def\eeq{\end{equation}}
\def\beqa{\begin{eqnarray}}
\def\eeqa{\end{eqnarray}}
\def\ba{\begin{eqnarray}}
\def\ea{\end{eqnarray}}
\def\bea{\begin{eqnarray}}
\def\eea{\end{eqnarray}}

\newcommand\as{\alpha_s}
\newcommand\f[2]{\frac{#1}{#2}}
\def\la{\lambda}

\def\beq{\begin{equation}}
\def\eeq{\end{equation}}
\def\beeq{\begin{eqnarray}}
\def\eeeq{\end{eqnarray}}
\def\to{\rightarrow}
\def\nn{\nonumber}

\def\b0{b_0}
\def\bone{b_1}
\def\btwo{b_2}

\def\th{\hat{\tau}}

\def\b0{b_0}
\def\bone{b_1}
\def\btwo{b_2}

\def\la{\lambda} 

\begin{document}

\begin{titlepage}
\renewcommand{\thefootnote}{\fnsymbol{footnote}}
\begin{flushright}
YITP-SB-14-44 \\
     \end{flushright}
\par \vspace{10mm}
\begin{center}
{\large \bf
Toward NNLL Threshold Resummation for Hadron Pair Production \\[4mm] 
in Hadronic Collisions}
\end{center}

\par \vspace{2mm}
\begin{center}
{\bf Patriz Hinderer${}^{\,a}$,}
\hskip .2cm
{\bf Felix Ringer${}^{\,a}$,}
\hskip .2cm
{\bf George F. Sterman${}^{\,b}$,}
\hskip .2cm
{\bf Werner Vogelsang${}^{\,a}$  }\\[5mm]
\vspace{5mm}
${}^{a}\,$ Institute for Theoretical Physics, T\"ubingen University, 
Auf der Morgenstelle 14, \\ 72076 T\"ubingen, Germany\\[2mm]
${}^{b}\,$ C.N.\ Yang Institute for Theoretical Physics,
Stony Brook University, Stony Brook, \\
New York 11794 -- 3840, U.S.A.\\
\end{center}

%%%%%%%%%%%%%%%%%%%%%%%%%%%%%%%%%%%%%%%%%%%%%%%%%%%%%%%%
%%%%%%%%%%%%%%%%%%%%%%%%%%%       ABSTRACT      %%%%%%%%%%%%%%%%%%%
%%%%%%%%%%%%%%%%%%%%%%%%%%%%%%%%%%%%%%%%%%%%%%%%%%%%%%%%

 \vspace{9mm}
\begin{center} {\large \bf Abstract} \end{center}
We investigate QCD threshold resummation effects beyond the next-to-leading logarithmic (NLL) 
order for the process $H_1 H_2\to h_1 h_2 X$ at high invariant mass of the produced hadron pair. 
We take into account the color structure of the underlying partonic hard-scattering cross sections
and determine the relevant hard and soft matrices in color space that contribute to the resummed
cross section at next-to-next-to-leading logarithmic (NNLL) accuracy. We present numerical results for 
fixed-target and collider regimes. We find a significant improvement  compared to previous 
results at NLL accuracy. In particular, the scale dependence of the 
resummed cross section is greatly reduced. Use of the most recent set of fragmentation functions also 
helps in improving the comparison with the experimental data. Our calculation provides a step towards a systematic 
NNLL extension of threshold resummation also for other hadronic processes, in particular for jet 
production.

\end{titlepage}  

\setcounter{footnote}{2}
\renewcommand{\thefootnote}{\fnsymbol{footnote}}

%%%%%%%%%%%%%%%%%%%%%%%%%%%%%%%%%%%%%%%%%%%%%%%%%%%%%%%%
%%%%%%%%%%%%%%%%%%%%%%%%%%%    INTRODUCTION     %%%%%%%%%%%%%%%%%%
%%%%%%%%%%%%%%%%%%%%%%%%%%%%%%%%%%%%%%%%%%%%%%%%%%%%%%%%

\section{Introduction}

The resummation of threshold logarithms in partonic hard-scattering cross sections 
contributing to hadronic scattering has received an ever-growing attention in recent years. 
On the one hand, resummation is phenomenologically relevant in many kinematical situations, 
ranging from fixed-target energies all the way to the LHC. At the same time, it offers insights
into the structure of perturbative corrections at higher orders, which among other things may 
provide benchmarks for explicit full fixed-order calculations in QCD. 

Threshold logarithms typically arise when the initial partons have just enough 
energy to produce the observed final state. In this case, the phase space available 
for gluon bremsstrahlung vanishes, resulting in large logarithmic corrections. Taking
the hadron-pair production cross section to be discussed in this paper as an example,
the partonic threshold is reached when $\hat{s}=\hat{m}^2$, 
that is, $\hat\tau\equiv\hat{m}^2/\hat{s}=1$, where $\sqrt{\hat{s}}$
is the partonic center-of-mass system (c.m.s.) energy and $\hat{m}$ the pair
mass of two outgoing produced partons that eventually fragment into the observed hadron pair.
The leading large contributions near threshold arise as $\as^k\left[ \ln^{2k-1}(1-\hat\tau)/
(1-\hat\tau)\right]_+$ at the $k$th order in perturbation theory, where $\as$ is the strong 
coupling and the ``plus'' distribution will be defined below. 
There is a double-logarithmic structure, with two powers of the logarithm arising
for every new order in the coupling. Subleading terms have fewer logarithms, so that
the threshold logarithms in the perturbative series take the general form 
\be\label{series1}
\sum_{k=0}^\infty\sum_{\ell=1}^{2k} \as^k \,{\cal A}_{k,\ell} \, \left(\f{ \ln^{2k-\ell}
(1-\hat\tau)}{1-\hat\tau}\right)_+\,,
\ee
with perturbative coefficients ${\cal A}_{k,\ell}$. One often refers to the all-order set of logarithms 
with a fixed $\ell$ as the $\ell$th {\it tower} of logarithms. As has been established in the
literature~\cite{dyresum,KS,BCMN}, threshold logarithms exponentiate after taking an
integral transform conjugate to the relevant kinematical variable ($\hat\tau$ in the above
example). Under this transform the threshold logarithms translate into logarithms of the 
transform variable $N$. The exponent may itself be written as a perturbative series and 
is only {\it single-logarithmic} in the transform variable. Ignoring for the moment the color 
structure of the underlying partonic cross section, the structure of the resummed
cross section becomes in transform space
\be\label{series2}
\left(1+\alpha_s C^{(1)}+\alpha_s^2 C^{(2)}+\ldots\right)\exp
\left[ \sum_{k=1}^\infty\sum_{\ell=1}^{k+1}{\cal B}_{k,\ell}\alpha_s^k\,\ln^\ell(N)\right]\;,
\ee
again with coefficients ${\cal B}_{k,\ell}$ and with ``matching coefficients'' $C^{(k)}$ that
ensure that at every fixed order the resummed cross section agrees with the exact fixed-order
one, up to corrections suppressed at threshold. They contain the full virtual corrections
at order $\alpha_s^k$, corresponding to contributions $\propto \delta(1-\hat\tau)$ in the
partonic cross section, and may be compared by comparison to a full fixed-order calculation
performed near threshold. Thanks to the exponentiated single-logarithmic
structure of the exponent, knowledge of the two leading towers $\as^k \ln^{k+1}(N)$ and 
$\as^k \ln^k(N)$, along with the coefficient $C^{(1)}$, is sufficient to predict the {\it three}
leading towers in the perturbative series~(\ref{series1}) for the cross section in $\hat\tau$-space. 
This is termed ``next-to-leading logarithmic'' (NLL) resummation. At full next-to-next-to-leading logarithmic 
(NNLL) accuracy, one needs three towers in the exponent and the two-loop coefficient $C^{(2)}$,
already providing control of five towers in the partonic cross section. 

While NLL resummation was the state of the art for many years, much progress has been made recently
on extending the framework to NNLL accuracy, or even beyond. The most advanced results have been
obtained for color-singlet processes such as Higgs production, where NNLL~\cite{Catani:2003zt,Ahrens:2008nc} and, 
most recently, even studies up to the N$^3$LL level~\cite{Bonvini:2014joa,Catani:2014uta} have been 
obtained, in which seven towers of 
logarithms are fully taken into account to all orders. This became possible when all threshold distributions 
at three-loop order were computed~\cite{Anastasiou:2014vaa}. For processes that are not characterized
by a color-singlet lowest-order (LO) hard scattering reaction, progress beyond NLL has also been made. 
For such processes, the resummation framework becomes more complex because the interference
between soft emissions by the various external partons in the hard scattering process becomes sensitive to 
the color structure of the hard scattering itself. This requires a color basis for the partonic 
scattering process which, as will be reviewed below, leads to a matrix structure of the soft 
emission~\cite{KS,BCMN,KOS,KO1}. This ultimately turns the exponential in~(\ref{series2}) into 
a sum of exponentials, each with its own set of matching coefficients $C^{(k)}$. An extensive list of 
color-non-singlet reactions of this type along with corresponding references to NLL studies may be 
found in~\cite{Catani:2013vaa}. Resummation studies beyond NLL have been presented in the context 
of top quark (pair) production~\cite{Kidonakis:2003qe,Beneke:2011mq,Czakon:2009zw,Yang:2014hya,Kidonakis:2010tc}, 
for single-inclusive hadron 
production~\cite{Catani:2013vaa}, and for squark and gluino production~\cite{Beneke:2010da,Beenakker:2013mva}.
At present, full NNLL resummation in the sense described above
is not yet possible for most processes, since the required two-loop matching coefficients are usually not
yet available (see, however, the recent calculation~\cite{Broggio:2014hoa} for massless
scattering). Nonetheless, with knowledge of the one-loop
matching coefficients an improvement of the resummation framework becomes possible already, providing
control of four (instead of five at full NNLL) towers in the partonic cross sections. A prerequisite
for this is that the appropriate color structure be taken into account for all ingredients in the resummed 
expression. 

In the present paper, we will develop such a partial NNLL resummation for the process of 
di-hadron production in hadronic collisions, collecting all necessary ingredients.
Previously, Ref.~\cite{Almeida:2009jt} presented a NLL study for this process which forms the basis for our paper. 
Kinematically, hadron pair production shares many features with the much simpler
color-singlet Drell-Yan process, if one confronts the produced partonic pair mass $\hat{m}$ 
with the invariant mass of the lepton pair. The interesting aspect of di-hadron production
is that it possesses all the color complexity of the underlying $2\to 2$ QCD hard scattering.
As such, the process becomes an ideal test for the study of QCD resummation beyond
NLL and can serve as a template for reactions of more significant phenomenological
interest, especially single or two-jet production in hadronic collisions. That said, di-hadron
production is phenomenologically relevant in its own right as experimental data
as a function of the pair's mass are available from various fixed-target 
experiments~\cite{na24,e711,e706}, as well as from the ISR~\cite{ccor}. In addition, 
di-hadron cross sections are also accessible at the Relativistic Heavy Ion Collider (RHIC).

Our paper is structured as follows. In Sec.~\ref{sec2} we recall the basic formulas for the di-hadron 
cross section as a function of pair mass at fixed order in perturbation theory, and display the role of the 
threshold region. In order for this paper to be self-contained, we recall a number of results from~\cite{Almeida:2009jt}. 
Section~\ref{sec3} presents details of the NNLL threshold resummation for the cross section. In particular, 
we derive the various hard and soft matrices in color space that are needed for the analysis. Here,
we make use of one-loop results available in the literature~\cite{Kunszt:1993sd,Bern:1990cu,Bern:1991aq} 
and compare to related work~\cite{Kelley:2010fn}. In Sec.~\ref{sec4} we give phenomenological results, 
comparing the threshold resummed calculations at NLL and NNLL to some of 
the available experimental data. Finally, we summarize our results in Sec.~\ref{sec5}.

\section{Hadron pair production near partonic threshold \label{sec2}}

\subsection{Perturbative cross section}

As in~\cite{Almeida:2009jt}, we consider the process $H_1(P_a) + H_2(P_b) \to h_1(P_c) +h_2(P_d) +X$
at measured pair invariant mass squared,
\beq
M^2 \equiv (P_c+P_d)^2 \; ,
\eeq
and at c.m.s. rapidities $\eta_1,\eta_2$ of the two produced hadrons. It is convenient to 
introduce
\beqa\label{baretadef1}
\Delta \eta &=& \frac{1}{2}(\eta_1 -\eta_2)  \; , \nonumber \\[1mm]
\bar{\eta} &=&\frac{1}{2}(\eta_1 +\eta_2) \; .
\eeqa
For sufficiently large $M^2$, the cross section for the process 
can be written in the factorized form
\beeq
M^4\frac{d \sigma^{H_1 H_2\to h_1 h_2 X}}{dM^2 d\Delta \eta  d\bar{\eta} }
&=&\sum_{abcd}
\int_0^1 dx_a dx_b  
dz_c  dz_d  \, f_a^{H_1}
(x_a,\mu_F)f_b^{H_2}(x_b,\mu_F)\,  z_c D_c^{h_1}(z_c,\mu_F)
z_dD_d^{h_2}(z_d,\mu_F) \nn \\
&& \hspace{10mm}
\times  \, \omega_{ab\to cd} \left(\th, \Delta \eta, 
\hat{\eta}, \as(\mu_R), \frac{\mu_R}{\hat{m}},\frac{\mu_F}{\hat{m}}\right)\;,
\label{taufac} 
\eeeq
where $\hat{{\eta}}$ is the average rapidity in the partonic c.m.s., which is related to $\bar\eta$ by
\beq
\hat{\eta}=\bar\eta-\frac{1}{2} \ln 
\left(\frac{x_a}{x_b}\right)  \; . 
\label{baretadef}
\eeq
The quantity $\Delta\eta$ is a difference of rapidities and hence 
boost invariant. The average and relative rapidities for the hadrons and 
their parent partons are the same, since all particles are taken to be massless.
The functions $f^{H_{1,2}}_{a,b}$ in Eq.~(\ref{taufac}) are the parton distribution functions
for partons $a,b$ in hadrons $H_{1,2}$ and $D_{c,d}^{h_{1,2}}$ the 
fragmentation functions for partons $c,d$ fragmenting into the 
observed hadrons $h_{1,2}$. The distribution functions are 
evaluated at a factorization scale $\mu_F$ that we choose to be the same
for the initial and the final state. $\mu_R$ denotes the renormalization scale,
which may differ from $\mu_F$.
The partonic momenta are given in terms of the hadronic ones by $p_a=x_a P_a$, 
$p_b=x_b P_b$, $p_c=P_c/z_c$, $p_d=P_d/z_d$.  We introduce
\beqa
S&=&(P_a+P_b)^2 \; , \nn \\[1mm]
\tau &\equiv& \frac{M^2}{S}  \; ,\nn \\[1mm]
\hat{s} &\equiv& \left( x_a P_a + x_b P_a \right)^2 = 
x_a x_b S\; ,\nn \\[1mm]
\hat{m}^2 &\equiv& \left( \frac{P_c}{z_c} + \frac{P_d}{z_d}\right)^2
= \frac{M^2}{z_cz_d} \; ,\nn \\[1mm]
\hat{\tau}&\equiv&\frac{\hat{m}^2}{\hat{s}} =
\frac{M^2}{x_ax_bz_cz_dS} = \frac{\tau}{x_ax_bz_cz_d} \; .
\eeqa
The $\omega_{ab\to cd}$ in Eq.~(\ref{taufac}) 
are the hard-scattering functions for the 
contributing partonic processes $ab\to cdX'$, where $X'$ denotes some additional 
unobserved partonic final state. Since the cross section in Eq.\ (\ref{taufac}) has 
been written in a dimensionless form, the $\omega_{ab\to cd}$ can be chosen to 
be functions of $\hat{m}^2/\hat{s}=\hat{\tau}$ and the ratios of $\hat{m}$ 
to the factorization and renormalization scales, as well as the 
rapidities and the strong coupling. They may be computed in QCD perturbation 
theory, where they are expanded as
\beq\label{overall}
\omega_{ab\to cd} = \left( \frac{\alpha_s}{\pi}\right)^2 \left[ 
\omega_{ab\to cd}^{\mathrm{LO}} + \frac{\alpha_s}{\pi}\,
\omega_{ab\to cd}^{\mathrm{NLO}} + \left(\frac{\as}{\pi} \right)^2
\omega_{ab\to cd}^{\mathrm{NNLO}} + \ldots \right] \; .
\eeq
Here we have separated the overall power of ${\cal{O}}(\as^2)$, 
which arises because the leading order partonic hard-scattering 
processes are the ordinary $2\to 2$ QCD scatterings.

\subsection{Threshold limit}

The limit $\hat{\tau} \to 1$ corresponds to the partonic threshold, where the 
hard-scattering uses all available energy to produce the pair. This is kinematically similar to 
the Drell-Yan process, if one thinks of the hadron pair replaced by
a lepton pair. The presence of fragmentation of course complicates the 
analysis somewhat, because only a fraction $z_cz_d$ of $\hat{m}^2$
is used for the invariant mass of the observed hadron pair. As shown in~\cite{Almeida:2009jt},
it is useful to introduce the variable 
\beqa
\tau' \equiv \frac{\hat{m}^2}{S}= \frac{M^2}{z_cz_dS}\; ,
\label{tauprime}
\eeqa
which may be viewed as the ``$\tau$-variable'' at the level of 
produced partons when fragmentation has not yet been taken 
into account, akin to the variable $\tau=Q^2/S$ in Drell-Yan. 

At LO, one has $\hat\tau=1$ and also $\hat{{\eta}}=0$. One can therefore write the LO term as 
\beq
\omega_{ab\to cd}^{\mathrm{LO}} \left(\th, \Delta \eta, 
\hat{{\eta}}\right)=
\delta\left( 1-\hat{\tau}\right)\, \delta \left(\hat{{\eta}}
\right)\,
\omega_{ab\to cd}^{(0)}(\Delta\eta) \; ,
\label{loomega}
\eeq
where $\omega_{ab\to cd}^{(0)}$ is a function of $\Delta \eta$
only. According to~(\ref{baretadef}), the second delta-function implies 
that $\bar\eta=\frac{1}{2} \ln(x_a/x_b)$.
At next-to-leading order (NLO), or overall ${\cal{O}}(\as^3)$, 
one can have $\hat{\tau}\neq 1$ and $\hat{{\eta}}\neq 0$. 
In general, as discussed in~\cite{Almeida:2009jt}, near partonic threshold 
the kinematics becomes ``LO like''. One has:
\ba
\omega_{ab\to cd} \left(\th, \Delta \eta, 
\hat{\eta}, \alpha_s(\mu_R), \frac{\mu_R}{\hat{m}},\frac{\mu_F}{\hat{m}} \right)
& = & \delta \left(\hat{{\eta}} \right)\,
\omega^{{\mathrm{sing}}}_{ab\to cd}\left(\th, \Delta \eta, 
\alpha_s(\mu_R),\frac{\mu_R}{\hat{m}},\frac{\mu_F}{\hat{m}} \right)\nn \\[2mm]
&+&\omega_{ab\to cd}^{{\mathrm{reg}}} \left(\th, \Delta \eta, 
\hat{\eta}, \alpha_s(\mu_R), \frac{\mu_R}{\hat{m}},\frac{\mu_F}{\hat{m}} \right)
\; , \label{allord}
\ea
where all singular behavior near threshold is contained in
the functions $\omega^{{\mathrm{sing}}}_{ab\to cd}$. Threshold 
resummation addresses this singular part to all orders in the 
strong coupling. All remaining contributions, which are subleading 
near threshold, are collected in the ``regular'' 
functions $\omega^{{\mathrm{reg}}}_{ab\to cd}$. Specifically,
for the NLO corrections, one finds the following structure:
\beeq
\omega_{ab\to cd}^{\mathrm{NLO}} \left(\th, \Delta \eta, 
\hat{\eta}, \frac{\mu_R}{\hat{m}},\frac{\mu_F}{\hat{m}} \right)
&=& \delta \left(\hat{{\eta}} \right)\,\left[ 
\omega^{(1,0)}_{ab\to cd}\left(\Delta \eta, \frac{\mu_R}{\hat{m}},\frac{\mu_F}{\hat{m}} \right)  \,
\delta (1- \hat{\tau}) \right. \nn \\[2mm]
&&+\left. 
\omega^{(1,1)}_{ab\to cd}\left(\Delta \eta, \frac{\mu_F}{\hat{m}}\right) 
\, \left( \frac{1}{1-\hat{\tau}}  \right)_+  
+ \omega^{(1,2)}_{ab\to cd}(\Delta \eta)
\left( \frac{ \log(1- \hat{\tau}) }{1-\hat{\tau}}  \right)_+  \; \right]
\nn \\[2mm]
&& +\, \omega^{{\mathrm{reg,NLO}}}_{ab\to cd}
\left(\hat{\tau},\Delta \eta,\hat{\eta}, \frac{\mu_R}{\hat{m}},\frac{\mu_F}{\hat{m}} \right)  \; , \label{NLO}
\eeeq
where the singular part near threshold is represented by the functions
$\omega_{ab\to cd}^{(1,0)}, \omega_{ab\to cd}^{(1,1)}, 
\omega_{ab\to cd}^{(1,2)}$, which are again functions
of only $\Delta\eta$, up to scale dependence. The ``plus''-distributions 
are defined by
\beq
\int_{x_0}^1 f(x)\left( g(x)\right)_+ 
dx\equiv \int_{x_0}^1 \left (f(x) -f(1) \right) \, 
g(x) dx - f(1) \int_0^{x_0} g(x) dx\; .
\eeq
The functions $\omega_{ab\to cd}^{(1,0)}, \omega_{ab\to cd}^{(1,1)}, \omega_{ab\to cd}^{(1,2)}$ 
were derived in~\cite{Almeida:2009jt} from an explicit NLO calculation near threshold. We will
use these results below as a useful check on the resummed formula and on the matching coefficients. 

\subsection{Mellin and Fourier transforms}

In order to prepare the resummation of threshold logarithms, we take integral 
transforms of the cross section. Following~\cite{Almeida:2009jt}, we first write the 
hadronic cross section in Eq.~(\ref{taufac}) as
\beq
M^4 \frac{d \sigma^{H_1 H_2\to h_1 h_2 X}}{dM^2 d\Delta \eta  d\bar{\eta} }
= \sum_{cd}\int_0^1 d z_c \, d z_d  \, z_c \,D_c^{h_1}(z_c,\mu_{F}) \,z_d\, 
D_d^{h_2}(z_d,\mu_{F}) \,
\Omega_{H_1 H_2 \to cd} \left( 
\tau', \Delta \eta, \bar{\eta},  \as(\mu_R), \frac{\mu_R}{\hat{m}},\frac{\mu_{{F}}}{\hat{m}}  
\right) \, ,
\label{Omegainsigma}
\eeq
where $\tau'=\hat{m}^2/S=\th x_ax_b$ and
\beeq
\Omega_{H_1 H_2\to cd}
\left( \tau', \Delta \eta, \bar{\eta},  \as(\mu_R), \frac{\mu_R}{\hat{m}} , \frac{\mu_F}{\hat{m}}  
\right) 
&\equiv&\sum_{ab}\int_0^1 d x_a \,d x_b \,
 f_a^{H_1} \left(x_a, \mu_F \right)\, f_b^{H_2}\left(x_b, \mu_F \right)  \nn
\\[2mm]
&&\times\;\;\omega_{ab\to cd} \left( \hat{\tau},  \Delta \eta,\hat{{\eta}},
\as(\mu_R),\frac{\mu_R}{\hat{m}}, \frac{\mu_F}{\hat{m}} 
\right) \; . \label{Dconv}
\eeeq
Taking Mellin moments of this function with respect to $\tau^{\prime}$ and a Fourier 
transform in $\bar{\eta}$, we obtain
\beqa
\label{Omemom}
&&\hspace*{-1.2cm}\int_{-\infty}^{\infty} 
d\bar\eta \, {\mathrm{e}}^{i \nu \bar{\eta}}\int_0^1 d\tau'\,
\left(\tau'\right)^{N-1} \Omega_{H_1 H_2\to cd}
\left( \tau', \Delta \eta, \bar{\eta},  \as(\mu_R),\frac{\mu_R}{\hat{m}}, \frac{\mu_F}{\hat{m}}   
\right) \nn \\[2mm]
&&\hspace*{-6mm}= \sum_{ab} \tilde{f}_a^{H_1}(N+1+i\nu/2,\mu_F)
\tilde{f}_b^{H_2}(N+1-i\nu/2,\mu_F)
\; \tilde{\omega}_{ab\to cd}
\left(N,\nu, \Delta \eta, \as(\mu_R),\frac{\mu_R}{\hat{m}}, \frac{\mu_F}{\hat{m}} 
\right)  \; , 
\eeqa
where $\tilde{f}_a^H(N,\mu_F)\equiv\int_0^1 x^{N-1}f_a^H(x,\mu_F)dx$, and 
\beq
\tilde{\omega}_{ab\to cd}
\left(N,\nu, \Delta \eta, \as(\mu_R),\frac{\mu_R}{\hat{m}}, \frac{\mu_F}{\hat{m}} 
\right)  \equiv \int_{-\infty}^{\infty}d\hat{\eta}\,
{\mathrm{e}}^{i \nu \hat{\eta}}\int_0^1 
d\hat\tau \,\hat\tau^{N-1} \, \omega_{ab\to cd}
\left(\hat{\tau}, \Delta \eta,\hat{\eta},\as(\mu_R),\frac{\mu_R}{\hat{m}}, \frac{\mu_F}{\hat{m}}  \right) \;.
\label{omegamom1}
\eeq
Near threshold, keeping only the singular  terms in~(\ref{allord}), the right-hand-side of this reduces to
\beq
\int_0^1 d\hat\tau \,\hat\tau^{N-1} \, 
\omega_{ab\to cd}^{\mathrm{sing}}
\left(\hat{\tau}, \Delta \eta,\alpha_s(\mu_R),
\frac{\mu_R}{\hat{m}},\frac{\mu_F}{\hat{m}} \right) \,\equiv\,
\tilde{\omega}_{ab\to cd}^{\mathrm{resum}}
\left(N,\Delta \eta, \as(\mu_R), \frac{\mu_R}{\hat{m}},\frac{\mu_F}{\hat{m}} \right)\;.
\label{omegamom2}
\eeq
We have labeled the new function on the right by the superscript ``resum'' as it is this quantity  
that contains all threshold logarithms and that threshold resummation addresses. As discussed 
in~\cite{Almeida:2009jt}, it is important here that we consider fixed $\hat m$ and fixed 
renormalization/factorization scales, which is achieved by isolating the fragmentation functions 
as in Eq.~(\ref{Omegainsigma}). Note that $\tilde{\omega}_{ab\to cd}^{\mathrm{resum}}$ 
depends on the Mellin variable $N$ only. All dependence on the Fourier variable $\nu$
resides in the moments of the parton distribution functions.

\section{Threshold resummation for hadron-pair production \label{sec3}}

In this section we present the framework for threshold resummation for di-hadron production 
at NNLL. We start by giving the main result and discussing its structure. Subsequently, we
will describe the various new ingredients in more detail.

\subsection{Resummation formula at next-to-next-to-leading logarithm \label{sec32}}

For di-hadron production near threshold, all gluon radiation is soft. Since all four
external partons in the hard scattering are ``observed'' in the sense that they
are either incoming or fragmenting partons, each of them makes the same
type of (double-logarithmic) contribution to the resummed cross section in moment 
space, given by a ``jet'' function $\Delta_i^N$ ($i=a,b,c,d$)  that takes
into account soft and collinear gluon radiation off an external 
parton~\cite{Almeida:2009jt,Cacciari:2001cw,Sterman:2006hu}. 
In addition, large-angle soft emission is sensitive to the color state of the hard scattering, 
giving rise to a trace structure in color space~\cite{KS,KOS}. The resummed partonic 
cross section in moment space then takes the following form~\cite{KS,BCMN,KOS,KO1,Almeida:2009jt}:
\beeq
\tilde{\omega}_{ab\to cd}^{\mathrm{resum}}
\left(N,\Delta \eta, \as(\mu_R), \frac{\mu_R}{\hat{m}},\frac{\mu_F}{\hat{m}} 
\right) &=& 
\Delta^{N+1}_a \left(\as(\mu_R), \frac{\mu_R}{\hat{m}},\frac{\mu_F}{\hat{m}} \right)
\Delta^{N+1}_b \left(\as(\mu_R), \frac{\mu_R}{\hat{m}},\frac{\mu_F}{\hat{m}} \right) \nn \\[2mm]
&\times& \Delta^{N+2}_c \left(\as(\mu_R), \frac{\mu_R}{\hat{m}},\frac{\mu_F}{\hat{m}} \right)
\Delta^{N+2}_d\left(\as(\mu_R), \frac{\mu_R}{\hat{m}},\frac{\mu_F}{\hat{m}} \right)\nn \\[2mm]
&\times&{\mathrm{Tr}} \left\{ H \left(\Delta\eta,\as(\mu_R) \right)\, {\cal{S}}^\dagger_N 
 \left(\Delta\eta,\as(\mu_R), \frac{\mu_R}{\hat{m}} \right) \,\right.\nn \\[2mm]
&& \left.\hspace*{7.5mm} S\left(\as(\hat{m}/\bar{N}),\Delta\eta \right)  {\cal{S}}_N  \left(\Delta\eta,\as(\mu_R), 
\frac{\mu_R}{\hat{m}} \right) \right\}_{ab \to cd}\nn \\[2mm]
& \times& \xi_R\left(\as(\mu_R), \frac{\mu_R}{\hat{m}} \right)\, \xi_F^{abcd}\left(\as(\mu_R), \frac{\mu_F}{\hat{m}}\right)\, \; .
\label{resumm}
\eeeq
This form is valid to all logarithmic order, up to corrections that are suppressed by powers of $1/N$, or $1-\th$. 
The additional functions $\xi_{R,F}$ do not contain threshold logarithms but are $N$-independent. They
serve to improve the dependence of the resummed cross section on the scales $\mu_R$ and $\mu_F$.
We will now discuss the various functions in Eq.~(\ref{resumm}) and their NNLL expansions. 

\subsubsection{Jet functions}

The radiative functions $\Delta_i^N$ are familiar from threshold resummation for the Drell-Yan process.
They exponentiate logarithms that arise due to soft-collinear gluon emission by the initial and final-state partons. 
In the $\overline{\rm{MS}}$ scheme, they are given by~\cite{dyresum,Catani:2003zt,Vogt:2000ci}
\bea
\Delta_i^N\left(\as(\mu_R), \frac{\mu_R}{\hat{m}},\frac{\mu_F}{\hat{m}} \right)
&=& R_i(\as(\mu_R))\, \exp\left\{\int_0^1 dz\,\f{z^{N-1}-1}{1-z}\right.\nn\\[2mm]
&&\hspace*{0cm}\times\, \left[ \int_{\mu_F^2}^{(1-z)^2 \hat{m}^2} \f{d\mu^2}{\mu^2} 
A_i(\as(\mu))+D_i(\as((1-z)\hat m))\right]\Bigg\}\, .
\label{Dfct}
\eea
The functions $A_i$ and $D_i$ may be calculated perturbatively as series in $\as$,
\bea
A_i(\as) & = & \frac{\as}{\pi} A_i^{(1)} +\left( \frac{\as}{\pi}\right)^2 A_i^{(2)} + 
\left(\frac{\as}{\pi}\right)^3 A_i^{(3)}+{\mathcal O}(\as^4) \nn \\[2mm]
D_i(\as) & = & \left(\f{\as}{\pi}\right)^2 D_i^{(2)}+{\mathcal O}(\as^3) \; ,
\eea
where, up to NNLL, one needs the coefficients~\cite{KT,Moch:2004pa,Catani:2001ic,Harlander:2001is,eric}
\beeq 
\label{A12coef} 
&&A_i^{(1)}= C_i
\;,\;\;\;\;\quad A_i^{(2)}=\frac{1}{2} \; C_i  \left[ 
C_A \left( \frac{67}{18} - \frac{\pi^2}{6} \right)  
- \frac{5}{9} N_f \right] \; , \nn \\[2mm] 
&&A_i^{(3)}=\f{1}{4}C_i\left[C_A^2 \left(\f{245}{24}-\f{67}{9}\zeta(2)+
\f{11}{6}\zeta(3)+\f{11}{5}\zeta(2)^2 \right)+C_F N_f\left(-\f{55}{24}+2\zeta(3) \right)\right. \nn \\[2mm]
&& \qquad\qquad\quad\;\, \left. +C_A N_f\left(-\f{209}{108}+\f{10}{9}\zeta(2)-\f{7}{3}\zeta(3) \right) -\f{1}{27}N_f^2 \right] \; , \nn \\[2mm]
&& D_i^{(2)}=C_i\left[C_A\left(-\f{101}{27}+\f{11}{3}\zeta(2)+\f{7}{2}\zeta(3)\right) + N_f\left(\f{14}{27}-\f{2}{3}\zeta(2)  \right) \right] \; ,
\eeeq 
with $N_f$ the number of flavors and 
\beeq\label{cqcg} 
&&C_q=C_F=\frac{N_c^2-1}{2N_c}=\frac{4}{3}  
\;, \;\;\;C_g=C_A=N_c=3 \; .
\eeeq
The $D_i$ term in the radiative factor $\Delta_i^N$ first appears at NNLL accuracy~\cite{dyresum,Catani:2003zt,Vogt:2000ci}. 
It takes into account logarithms that arise from soft gluons that are emitted at large angles. Incoming and outgoing
external lines of a given parton type carry the same $D_i$ term, as discussed in the~Appendix. 

Finally, the coefficient $R_i$ in Eq.~(\ref{Dfct}) ensures that our soft functions for this process are defined
relative to that for the Drell-Yan process; again see the~Appendix for details. To the order we consider, we have
\be \label{Ri}
R_i(\as)\,=\,1-\frac{3\as}{4\pi}\,A_i^{(1)} \,\zeta(2)+{\cal O}(\as^2)\;.
\ee

Evaluating the integrals in Eq.~(\ref{Dfct}), one obtains an explicit expression for the NNLL expansion of the function 
$\Delta_i^N$:
\beeq\label{DfctNLL}
\Delta_i^N\left(\as(\mu_R), \frac{\mu_R}{\hat{m}}, \frac{\mu_F}{\hat{m}} \right)
& = & \tilde{R}_i(\as(\mu_R))\, \exp\left\{
h_i^{(1)} (\lambda)\,\ln \bar{N} + h_i^{(2)} \left(\lambda, \frac{\mu_R}{\hat{m}}, \frac{\mu_F}{\hat{m}} \right)
\right. \nn \\[2mm]
&&\hspace*{3.1cm}+\,\left. \as(\mu_R)\,  h_i^{(3)} \left(\lambda, \frac{\mu_R}{\hat{m}}, 
\frac{\mu_F}{\hat{m}} \right)\right\}\; .
\eeeq
Here $\tilde{R}_i$ is a combination of $R_i$ in Eq.~(\ref{Ri}) and a $\pi^2$-term arising in the NNLL 
expansion~\cite{Catani:2003zt}:  
\be \label{Ritilde}
\tilde{R}_i(\as)\,=\,1+\frac{\as}{4\pi}\,A_i^{(1)} \,\zeta(2)+{\cal O}(\as^2)\;.
\ee
In~(\ref{DfctNLL}) we have furthermore defined $\lambda = b_0 \as (\mu_R) \ln(N{\mathrm{e}}^{\gamma_E})$ with 
$\gamma_E$ the Euler constant. In the following we denote $N{\mathrm{e}}^{\gamma_E}\equiv \bar{N}$. 
The functions $h_i^{(1)},h_i^{(2)},h_i^{(3)}$ read
\beeq
h_i^{(1)}(\lambda) &=&\f{A_i^{(1)}}{ 2 \pi \b0 \lambda}
\left( 2 \lambda + (1-2\lambda)\ln ( 1- 2 \lambda) \right) \; , \\[1mm]
h_i^{(2)}\left(\lambda,\frac{\mu_R}{\hat m},\frac{\mu_F}{\hat m}\right) &= & -
\frac{A_i^{(2)}}{2 \pi^2 b_0^2} \left[2\lambda + \ln (1-2\lambda)\right] \nn\\[2mm]
&+&\frac{A_q^{(1)}b_1}{2\pi b_0^3}\left[2\lambda + 
\ln(1-2\lambda)+ \frac{1}{2} \ln^2(1-2\lambda)\right] \nonumber  \\[2mm]
& -& \frac{A_i^{(1)}}{2\pi b_0} \left[2 \lambda + 
\ln(1-2\lambda)\right] \ln\frac{\mu_R^2}{\hat m^2}  + \frac{A_i^{(1)}}{\pi b_0} \lambda
 \ln \frac{\mu_F^2}{\hat m^2}\; ,
\eeeq
and~\cite{Vogt:2000ci}
\beeq
h_i^{(3)}\left(\lambda,\frac{\mu_R}{\hat m},\frac{\mu_F}{\hat m}\right) &= & 
 \f{2 A_i^{(1)}}{\pi} \zeta(2) \f{\la}{1-2\la}- \f{A_i^{(2)} \bone }{2\pi^2 \b0^3} \f{1}{1-2\la}\left[2\la+\ln(1-2\la) +2 \la^2 \right]
\nn \\[2mm] 
& +& \f{A_i^{(1)}  \bone^2}{2\pi \b0^4 (1-2\la)} 
\left[2\la^2+ 2\la \ln(1-2\la)+\f{1}{2} \ln^2(1-2\la) \right] \nn \\[2mm] 
& +& \f{A_i^{(1)}  \btwo}{2\pi\b0^3} 
\left[2\la+\ln(1-2\la) +\f{2\la^2}{1-2\la} \right] 
+ \f{A_i^{(3)}}{\pi^3 \b0^2} \f{\la^2}{1-2\la} \nn \\[2mm] 
&+& \f{A_i^{(2)}}{\pi^2 \b0} \,\la \ln\f{\mu^2_F}{\hat m^2} 
- \f{A_i^{(1)}}{2\pi} \, \la  \ln^2\f{\mu^2_F}{\hat m^2} 
+ \f{A_i^{(1)}}{\pi} \la \ln\f{\mu_R^2}{\hat m^2} \ln\f{\mu^2_F}{\hat m^2} \nn \\[2mm] 
&-& \f{1}{1-2\la} \Big( 
 \f{ A_i^{(1)} \bone}{2\pi \b0^2} \left[ 2\la+\ln(1-2\la) \right]
 - \f{2 A_i^{(2)}}{\pi^2 \b0}  \la^2 \Big)
\ln\f{\mu_R^2}{\hat m^2} \nn \\[2mm]
&+ &\f{A_i^{(1)}}{\pi} \f{\la^2}{1-2\la}\ln^2\f{\mu_R^2}{\hat m^2} 
- \f{ D_i^{(2)}}{2\pi^2 \b0} \f{\la}{1-2\la} \; .
\label{EE}
\eeeq
Here $b_0,\,b_1,\,b_2$ are the first three coefficients of the QCD beta function which are given by~\cite{Tarasov:1980au,Larin:1993tp}
\bea\label{thebs}
b_0 & = & \f{1}{12\pi} \left(11C_A-2 N_f\right)\; , \qquad b_1 = \f{1}{24\pi^2}\left(17C_A^2-5C_AN_f-3C_F N_f\right) \; , \nn \\[2mm]
b_2 & = & \f{1}{64\pi^3}\left(\f{2857}{54} C_A^3- \f{1415}{54} C_A^2 N_f-\f{205}{18} C_A C_F N_f+
\f{78}{54} C_A N_f^2+\f{11}{9} C_F N_f^2\right) \; .
\eea

\subsubsection{Color trace contribution}

Next we discuss the trace ${\mathrm{Tr}} \{ H {\cal{S}}^\dagger_N S  {\cal{S}}_N\}$ in color space in Eq.~(\ref{resumm}). 
We note that this is the only contribution to the resummed cross section that depends on the 
difference of the rapidities $\Delta\eta$. Each of the factors $H_{ab\to cd}$, ${\cal S}_{N,ab\to cd}$, $S_{ab\to cd}$ 
is a matrix in the space of color exchange operators~\cite{KS,KOS}. The $H_{ab\to cd}$ are the hard-scattering functions. 
They are perturbative and have the expansions
\beq\label{eq:Hexpansion}
H_{ab\to cd}\left(\Delta\eta,\as \right)= \left(\f{\as}{\pi} \right)^2\left[H_{ab\to cd}^{(0)}\left(\Delta\eta\right)+
\frac{\alpha_s}{\pi} H_{ab\to cd}^{(1)}\left(\Delta\eta \right)+ {\cal O}(\alpha_s^2)\right] \; .
\eeq
The LO (i.e. ${\cal O}(\as^2)$) contributions $H_{ab\to cd}^{(0)}$ may be found in~\cite{KS,KOS,KO1}. For 
resummation beyond NLL accuracy, one needs all entries of the NLO hard-scattering matrices $H^{(1)}_{ab\to cd}$. 
These matrices may be extracted from a color decomposed one-loop calculation~\cite{Almeida:2009jt,Kelley:2010fn}. 
We will outline the derivation of the first-order corrections $H^{(1)}_{ab\to cd}$ in Section~\ref{HSF}. We note that
they depend in principle also on the renormalization scale $\mu_R$, in the form of a term $\propto
\ln(\mu_R/\hat{m})H_{ab\to cd}^{(0)}$. This dependence, however, has been absorbed
into the contribution involving the function $\xi_R$ in~(\ref{resumm}); see below.

The $S_{ab\to cd}$ are known as soft functions. In general, they depend on the rapidity difference $\Delta\eta$ and 
on the strong coupling whose argument is to be set to $\hat m/\bar{N}$~\cite{KS,KOS,Czakon:2009zw}. This
dependence on $\hat m/\bar N$ and hence on $N$ occurs first at NNLL. The soft functions have the expansion 
\beq\label{eq:soft1}
S_{ab\to cd}\left(\as(\hat{m}/\bar{N}),\Delta\eta \right)= S_{ab\to cd}^{(0)}+ \frac{\alpha_s(\hat m/\bar N)}{\pi}
S_{ab\to cd}^{(1)}\left(\Delta\eta\right)+ {\cal O}(\alpha_s^2) \; .
\eeq
Relating the coupling at scale $\hat m/\bar N$ to that at scale $\mu_R$, one can construct the explicit $N$-dependence of the 
soft matrix at NLO. To the accuracy of resummation that we are considering in this work, it is sufficient to use
\be
\alpha_s(\hat m/\bar N)=\f{\as(\mu_R)}{1-2\lambda}\; .
\ee
The LO expressions $S_{ab\to cd}^{(0)}$, which are independent of $\Delta\eta$, may also be found in~\cite{KS,KOS,KO1}. 
Like the hard-scattering matrices $H^{(1)}_{ab\to cd}$, at NNLL accuracy, we need the explicit expressions for the 
full NLO soft-matrices $S^{(1)}_{ab\to cd}$. These may be extracted by performing a color-decomposed calculation 
of the $2\rightarrow 3$ contributions to the partonic cross sections in the eikonal approximation, as will be described in
Section~\ref{SF}. 

The resummation of wide-angle soft gluons is contained in ${\cal{S}}_{ab\to cd}$. The two exponentials 
${\cal{S}}^\dagger_N$ and ${\cal{S}}_N$ that enclose the soft function $S_{ab\to cd}$ within the 
trace structure appear when solving the renormalization group equation for the soft function~\cite{KS,KOS,KO1}. 
The exponentials are given in terms of soft anomalous dimensions $\Gamma_{ab\to cd}$:
\beeq
\label{GammaSoft}
{\cal S}_{N,ab\to cd} \left(\Delta\eta,\as(\mu_R), \frac{\mu_R}{\hat{m}} 
\right) &=& {\cal P}\exp\left[ 
\frac{1}{2} \int^{\hat{m}^2/\bar{N}^2}_{\hat{m}^2} 
\frac{d\mu^2}{\mu^2} \Gamma_{ab\to cd} 
\left(\Delta\eta,\as(\mu) \right)\right] \, ,
\eeeq
where ${\cal P}$ denotes path ordering. The soft anomalous dimension matrices start at ${\cal O}(\alpha_s)$,
\beq\label{eq:Gamma12}
\Gamma_{ab\to cd}\left(\as,\Delta\eta\right)=\frac{\as}{\pi}\,\Gamma^{(1)}_{ab\to cd}
\left(\Delta\eta\right)+ \left(\frac{\as}{\pi}\right)^2\, \Gamma^{(2)}_{ab\to cd}(\Delta\eta) + {\cal O}(\as^3) \; .
\eeq
Their first-order terms are presented in~\cite{KS,KOS,KO1,msj}. We will discuss the $\Gamma_{ab\to cd}$ 
matrices in more detail in Section~\ref{SF}. For NNLL resummation, we also need to take into account 
the second-order contributions $\Gamma^{(2)}_{ab\to cd}$ which were derived in~\cite{twoloopad} and
are determined by the one-loop terms:
\be
\Gamma^{(2)}_{ab\to cd}(\Delta\eta) = \frac{K}{2} \Gamma^{(1)}_{ab\to cd}(\Delta\eta)\; ,
\ee
where $K=C_A(67/18-\pi^2/6)-5 N_f/9$. We also give here our result for the NNLL 
expansion of the integral in Eq.~(\ref{GammaSoft}):
\ba
\label{GammaSoft1}
\ln {\cal S}_{N,ab\to cd} \left(\Delta\eta,\as, \frac{\mu_R}{\hat{m}} \right)&=& \Gamma^{(1)}_{ab\to cd}
\left(\Delta\eta\right)\left[ \f{\ln(1 - 2 \lambda)}{2\pi b_0}+\f{\as}{\pi}\f{1}{2b_0^2\pi(1-2\lambda)} \right.\nn \\[2mm]
&\times&\left.
\bigg(b_1 \pi(2\lambda+\ln(1-2\lambda))\bigg. \left.  -b_0\lambda\left(K+2\pi b_0\ln\f{\mu_R^2}{\hat m^2}\right)\right) \right] \; .
\ea
We note that in our phenomenological applications we follow~\cite{Almeida:2009jt} and perform the 
exponentiation of the matrices numerically by iterating the exponential series to an adequately high order.

In order to clarify the roles of the various matrices appearing in the color trace,
it is instructive to analyze the structure of the resummed cross section~(\ref{resumm}) in Mellin space 
after expansion to NLO:
\beeq
\tilde{\omega}_{ab\to cd}^{\mathrm{resum}}
\left(N,\Delta \eta, \as(\mu_R), \frac{\mu_R}{\hat{m}} ,\frac{\mu_F}{\hat{m}} 
\right)&=& \left(\f{\as(\mu_R)}{\pi}\right)^2 \left[ {\mathrm{Tr}}\{H^{(0)} S^{(0)}\}_{ab\to cd}\,\left\{ 1 + 2 b_0 
\alpha_s(\mu_R) \ln\frac{\mu_R^2}{\hat{m}^2} \right.\right.\nn\\[2mm]
&&\hspace*{-6.2cm}\,+\,\left.\frac{\alpha_s(\mu_R)}{\pi}\sum_{i=a,b,c,d}\left(
A_i^{(1)}\ln^2\bar{N} +\frac{1}{4}A_i^{(1)}\zeta(2)+ \left(A_i^{(1)}\ln\bar{N} + \frac{1}{2}B_i^{(1)}\right)
\ln\frac{\mu_F^2}{\hat{m}^2}\right)\right\}\nn \\[2mm]
&&\left.\hspace*{-6.2cm}\,+\,\frac{\alpha_s(\mu_R)}{\pi}\,{\mathrm{Tr}}\Big\{ -\left[ 
H^{(0)} (\Gamma^{(1)})^\dagger  S^{(0)}+ H^{(0)} S^{(0)}\Gamma^{(1)} 
\right]\ln \bar{N}\,+\,H^{(1)} S^{(0)}+ H^{(0)} S^{(1)}\Big\}_{ab\to cd}\right]
+{\cal O}(\alpha_s^4)\; .\nn\\
\label{nloexp}
\eeeq 
Here the term $\propto \zeta(2)$ arises from the coefficient $\tilde{R}_i$ in~(\ref{Ritilde}). 
We have anticipated the contributions by the functions $\xi_R$ and $\xi_F$ in~(\ref{resumm})
that will be specified in the next subsection. $\xi_R$ yields the term involving the renormalization scale, 
and $\xi_F$ contributes the ones $\propto B_i^{(1)}$, with
\be\label{bqbg}
B_q^{(1)}=-\f{3}{2}C_F\; , \qquad   B_g^{(1)} = -2\pi b_0 \; .
\ee

The term ${\mathrm{Tr}}\{H^{(0)} S^{(0)}\}_{ab\to cd}$ in~(\ref{nloexp}) is proportional to the 
LO function $\omega_{ab\to cd}^{(0)}(\Delta\eta)$ introduced in Eq.~(\ref{loomega}). 
In~\cite{Almeida:2009jt} (as in many previous studies of threshold resummation for 
hadronic hard-scattering), the combination $\mathrm{Tr}[H^{(1)} S^{(0)}+ H^{(0)} S^{(1)}]$, which carries no 
dependence on $N$, was extracted as a whole by matching the expression in Eq.~(\ref{nloexp}) to the NLO calculation 
at threshold. Of course, this is not sufficient for determining the full first-order matrices $H^{(1)}$ and $S^{(1)}$. 
However, it is a valid approach at NLL accuracy, where the three most dominant towers of logarithms are 
taken into account. For a given fixed-order expansion to ${\cal O}(\as^k)$, the following terms are under control:
\be
\as^k\; \left\{ \ln^{2k}\bar N,\;\; \ln^{2k-1}\bar N,\;\; \ln^{2k-2}\bar N\right\} \; .
\ee
It is straightforward to see that the hard and soft matrices will contribute to the third tower of threshold logarithms 
always in the combination ${\mathrm{Tr}}\{H^{(1)}S^{(0)}+H^{(0)}S^{(1)}\}_{ab\to cd}$ in the following way:
\be\label{TrSH1}
\left(\f{\as}{\pi}\right)^k\, \f{\sum_{i} A_i^{(1)}}{(k-1)!} \, \mathrm{Tr}\left\{H^{(1)} S^{(0)}+ H^{(0)} S^{(1)}\right\}_{ab\to cd}\, \ln^{2k-2}\bar N \; .
\ee
Hence, to NLL, it is sufficient to know the combined expression of $H^{(1)}$ and $S^{(1)}$ instead of having to compute the full matrices 
separately. It is then legitimate to that order to approximate the trace term in the resummed formula by
\beq\label{Ccoeff1}
{\mathrm{Tr}} \left\{ H {\cal{S}}_N^\dagger
S  {\cal{S}}_N \right\}_{ab \to cd,\,{\rm NLL}} \,=\,
\left(1+\frac{\alpha_s}{\pi}\,C^{(1),\,{\rm NLL}}_{ab \to cd}\right)\,
{\mathrm{Tr}} \left\{ H^{(0)} {\cal{S}}_N^\dagger 
S^{(0)} {\cal{S}}_N \right\}_{ab \to cd} \; ,
\eeq
where
\beq
C^{(1),\,{\rm NLL}}_{ab \to cd}\left( \Delta\eta\right)
\equiv \frac{{\mathrm{Tr}} \left\{ 
H^{(1)} S^{(0)}+ H^{(0)} S^{(1)}\right\}_{ab\to cd}}{{\mathrm{Tr}} \left\{ 
H^{(0)} S^{(0)}\right\}_{ab\to cd}}\;.
\label{Ccoeff}
\eeq
This was the approach adopted in~\cite{Almeida:2009jt} and also, for example, in studies on single-inclusive hadron~\cite{ddfwv} 
or jet production~\cite{deFlorian:2007fv}. 

On the other hand, in order to control the fourth tower of logarithms, $\as^k\,\ln^{2k-3}\bar N$, one needs to 
know $H^{(1)}$ and $S^{(1)}$ explicitly as they also appear separately in various combinations with the anomalous
dimension matrices. Computation of the full matrices is therefore a necessary ingredient for NNLL resummation. 
Clearly, having the matrices at hand, one can compute also the known combination 
${\mathrm{Tr}}\{H^{(1)}S^{(0)}+H^{(0)}S^{(1)}\}_{ab\to cd}$,
which provides an important cross-check on them. We stress further that, in order to fully take into account also 
the fifth tower $\as^k\,\ln^{2k-4}\bar N$ at NNLL, one would need to know the full matrices $H^{(2)}$ and $S^{(2)}$ and 
perform a matching to NNLO.
Although  $H^{(2)}$ became available very recently~\cite{Broggio:2014hoa},  this is beyond the scope of the present work.

We finally note that a new feature which first appears at NNLL is that the hard-scattering matrix $H$ obtains an imaginary 
part. This is because $H$ is constructed from virtual corrections to partonic $2\to 2$ scattering, 
which contain logarithms of ratios of space- and timelike invariants. We write
\be
H = H_R+iH_I
\ee
with $H_R$ and $H_I$ real. It turns out that $H_R$ is a symmetric matrix, whereas $H_I$ is antisymmetric; 
see Section~\ref{HSF}. Hence, the hard-scattering matrix $H$ as a whole is hermitian, as it should be. 
The imaginary part $H_I$ contributes to the resummed cross section due to the fact that the remaining 
terms inside the color trace in the resummed cross section~(\ref{resumm}), 
$M\equiv {\mathrm{e}}^{\Gamma^\dagger}S{\mathrm{e}}^{\Gamma}$, also develop an imaginary part since the anomalous
dimension matrices are complex-valued~\cite{KS,KOS}. $M$ is also hermitian as the soft matrix $S$ is symmetric,
and therefore we may also decompose $M=M_R+iM_I$ with $M_R$ symmetric and $M_I$ antisymmetric. 
The trace $\mathrm{Tr}\left\{HM\right\}$ is then real, as it must be, but both the real and imaginary parts of $H,M$ 
contribute:
\be
\mathrm{Tr}\left\{HM\right\}=\mathrm{Tr}\left\{H_R M_R\right\}-\mathrm{Tr}\left\{H_I M_I\right\} \;.
\ee
Note that the contribution by the imaginary part of $H$ drops out from $\mathrm{Tr}[H^{(1)} S^{(0)}+ H^{(0)} S^{(1)}]$,
so that it is not present at NLL. Performing an analytical fixed-order expansion of our NNLL resummed result, 
we find that the imaginary parts of $H$ and $M$ first start to play a role at $\mathrm{N^3LO}$, where they 
contribute to the fifth tower, $\as^3\,\ln^2\bar N$. We note, however, that the imaginary parts of $\Gamma_{ab\to cd}$
also contribute to the real part of $M$, since $M={\mathrm{e}}^{\Gamma^\dagger}S{\mathrm{e}}^{\Gamma}$.
It turns out that they already appear in the {\it fourth} tower of logarithms. In this way we see that the imaginary parts 
of the various contributions are important ingredients of the NNLL resummed cross section.

\subsubsection{Functions $\xi_R$ and $\xi_F$}

The $N$-independent function $\xi_F^{abcd}$ in Eq.~(\ref{resumm})
addresses the factorization scale dependence of the 
cross section~\cite{KS,KOS,KO1,Laenen:2000ij}:
\be\label{xi22}
\ln \xi_F^{abcd}\left(\as(\mu_R),\f{\mu_F}{\hat m} \right) = -\f{1}{2}\sum_{i=a,b,c,d}
\int_{\mu_F^2}^{\hat m^2}\f{d\mu^2}{\mu^2} \f{\as(\mu)}{\pi} B_i^{(1)} \; ,
\ee
where we are summing over all four external partons. The coefficients $B_i^{(1)}$, which have 
been given in Eq.~(\ref{bqbg}), correspond to the $\delta$-function contributions to the corresponding 
LO diagonal parton-to-parton splitting functions and thus depend on whether the considered parton $i$ 
is a quark or a gluon. As follows from~\cite{KSV}, the function $\xi_F^{abcd}$ takes into account all $N$-independent 
pieces corresponding to the evolution of parton distributions and fragmentation functions between
scales $\mu_F$ and $\hat{m}$. Again its first-order contribution would explicitly appear in 
$H_{ab\to cd}^{(1)}$, from where it has been absorbed. It is straightforward to expand~(\ref{xi22}) 
to the desired NNLL accuracy. 

$\xi_R$ governs the renormalization scale dependence of the resummed cross section. This function was
also introduced in~\cite{KO1}. $\xi_R$ essentially serves to set the scale in the strong coupling 
constant in the overall factor $\alpha_s^2$ (see Eq.~(\ref{overall})) of the cross section to $\hat m$:
\be\label{xi11}
\ln \xi_R\left(\as(\mu_R),\f{\mu_R}{\hat m} \right) = 2 \int_{\mu_R^2}^{\hat m^2}\f{d\mu^2}{\mu^2}
\beta(\as(\mu))\; .
\ee
Evaluating the integral while keeping only the first two terms in the QCD $\beta$-function,
\beq
\beta(\as) \equiv \frac{1}{\as} \frac{d\as}{d\log(\mu^2)}=-b_0\alpha_s-b_1\alpha_s^2+{\cal O}(\alpha_s^3)\,,
\eeq
and expanding the result up to second order in $\as$, we find
\be\label{xi111}
\ln \xi_R\left(\as,\f{\mu_R}{\hat m} \right) = 2 b_0\as \ln\f{\mu_R^2}{\hat m^2}+\as^2
\left(2b_1 \ln\f{\mu_R^2}{\hat m^2}+b_0^2 \ln^2\f{\mu_R^2}{\hat m^2} \right) \; .
\ee
Here $b_0,b_1$ are as given in~(\ref{thebs}). The first term on the right 
reproduces the explicit $\mu_R$-dependence of the first-order hard-scattering function that we have 
chosen to pull out of $H_{ab\to cd}^{(1)}$. The additional terms generated by this expression produce
higher-order scale-dependent contributions that will occur in the perturbative series. When combined 
with resummation at NNLL level, they necessarily help to stabilize the cross section with respect to 
changes in $\mu_R$, as we shall discuss in more detail now.

Following~\cite{Glover:2002gz} and suppressing all arguments except for the renormalization scale, 
we write the perturbative expansion of a generic partonic cross section $\omega$ as
\be
\omega = \sum_{k=0}^\infty \as^{k+2}(\mu_R)\,\omega^{(k)}(\mu_R)\;.
\ee
The LO coefficient $\omega^{(0)}$ is independent of $\mu_R$; all higher-order terms depend
on $\mu_R$ through the logarithm $L\equiv \ln(\mu_R^2/\hat m^2)$.
Truncating the series at some fixed $k=n$, the uncertainty introduced by the  renormalization scale 
dependence is of the order of ${\cal O}(\as(\mu_R)^{n+3})$. In the following we consider as an example
the renormalization scale dependence after truncation to next-to-next-to-leading order (NNLO), which is given by
\ba\label{eq:scale1}
\omega|_{{\mathrm{NNLO}}} & = & \as^2(\mu_R)\, \omega^{(0)}+\as^3(\mu_R)\, \omega^{(1)}(\mu_R)+\as^4(\mu_R)\, 
\omega^{(2)}(\mu_R)\nn\\[2mm]
&=&\as^2(\mu_R)\,  \omega^{(0)} + \as^3(\mu_R)\, \left( \omega'^{\,(1)}+2b_0\, L\,   \omega^{(0)}\right) \nn \\[2mm]
& + & \as^4(\mu_R)\, \left( \omega'^{\,(2)}+3b_0\, L\,   \omega'^{\,(1)}+ (3 b_0^2\, L^2+2 b_1\, L)\,  \omega^{(0)}\right) \; ,
\ea
where the coefficients $\omega'^{\,(k)}$ denote the terms in $\omega^{(k)}$ that do not carry any dependence on $\mu_R$. 
As is well-known, the $\mu_R$-dependence of the NNLO cross section is entirely determined by the NLO terms in the 
perturbative expansion. 

We may now compare the general expression in Eq.~(\ref{eq:scale1}) to an NNLO expansion 
of the resummed cross section $\tilde\omega^{\mathrm{resum}}$ at either NLL or NNLL. 
First of all, we find that the NLO scale dependence and the contribution $(3 b_0^2\, L^2+2 b_1\, L)\, \omega^{(0)}$ 
at NNLO are entirely reproduced by the exponential $\xi_R$ in Eqs.~(\ref{xi11}) and~(\ref{xi111}). The 
interesting term at NNLO is now the term $3b_0\,L\, \omega'^{\,(1)}$ in the last line. Out of the five towers of 
threshold logarithms that appear at NNLO, the renormalization scale dependence resides only in the lowest three. 
Indeed, as can be seen from the explicit NLO expansion given in Eq.~(\ref{nloexp}), the coefficient $\omega'^{\,(1)}$ 
contains terms proportional to $\ln^2\bar N,\,\ln\bar N,\,1$ which, at NNLO, correspond to the $3^{\mathrm{rd}}$, 
$4^{\mathrm{th}}$ and $5^{\mathrm{th}}$ towers. If we now compare to the NNLO expansion of the {\it NLL}-resummed
cross section, we find that only the scale dependence of the $3^{\mathrm{rd}}$ tower is correctly reproduced. 
For the $4^{\mathrm{th}}$ and $5^{\mathrm{th}}$ tower, that are not fully taken into account at NLL, we find a 
factor of $2b_0\, L$ instead of $3b_0\, L$ multiplying the corresponding part of the coefficient $\omega'^{\,(1)}$. 
If instead resummation is performed at NNLL, the scale dependence in the $4^{\mathrm{th}}$ tower is correctly 
reproduced as well, whereas in the $5^{\mathrm{th}}$ tower the incorrect factor $2b_0\, L$ remains. 
(In addition, of course, the scale-independent coefficient $\omega'^{\,(2)}$ also changes). As it turns out, 
going from NLL to NNLL leads to a dramatic reduction of the renormalization scale uncertainty
of the resummed cross section, as will be seen in our numerical studies in Sec.~\ref{sec4}. 

\subsection{Hard-scattering function\label{HSF}}

In this Section we present our derivation of the matrices $H_{ab\to cd}^{(1)}$. We note that these
were also determined in~\cite{Kelley:2010fn}; the results of our independent computation are in agreement 
with that reference. As the resulting expressions become rather lengthy in general, we present explicit results 
only for the simplest partonic channel, $qq'\rightarrow qq'$. For ease of notation, we will usually drop
the ubiquitous subscript ``$qq'\to qq'\,$'' of the matrices. We also refer the reader to Ref.~\cite{KO1}, where
many details of the relevant color bases have been collected. In fact, for each partonic channel we adopt the
corresponding color basis from that reference. We note that our choice differs from the one in~\cite{Kelley:2010fn}, 
where an overcomplete basis was chosen for the $gg\rightarrow gg$ channel.

\subsubsection{Color basis and lowest-order contribution}

We consider the partonic process
\beq
q(p_1,a)q'(p_2,b) \rightarrow q(p_3,c)q'(p_4,d)\;,
\eeq
where the $p_i$ are the momenta of the incoming and outgoing quarks, and the indices $a$, $b$, $c$, $d$ denote their 
color. Given the fact that the leading-order process has only a $t$-channel contribution, it is convenient to 
choose the $t$-channel octet-singlet color basis which leads to a simple form for the lowest-order hard-scattering 
matrix $H^{(0)}$. The contributing color tensors in this basis are given by (1=octet, 2=singlet)
\ba
{\cal C}_{1}&\equiv&T^{g}_{ca}T^{g}_{db}\,=\,
\frac{1}{2}\left(\delta_{ad}\,\delta_{bc}-\frac{1}{N_c}\,\delta_{ac}\,\delta_{bd}\right)\;,\nn \\[2mm]
{\cal C}_{2}&\equiv&\delta_{ac}\delta_{bd}\;,
\ea
where $T^g$ is the generator in the fundamental representation and the indices $a,b,c,d$ will 
be kept implicit throughout most of our discussion. The soft and hard functions 
become matrices in this basis, whose entries are determined as the coefficients multiplying
the respective tensor structures. The elements of the leading-order contribution to the soft 
function $S^{(0)}$ in Eq.~(\ref{eq:soft1}) are given by ($I,J=1,2$)
\beq\label{eq:S0-trace}
(S^{(0)})_{JI}\equiv{\mathrm{Tr}}[{\cal C}_J^\dagger {\cal C}_I]\equiv
\sum_{a,b,c,d=1}^{N_c}{\cal C}_J^* {\cal C}_I\;.
\eeq
In our basis one finds
\beq\label{S0mat}
S^{(0)}=
\begin{pmatrix}
\frac{N_{c}^{2}-1}{4} & 0 \\[1mm] 0 & N_{c}^2
\end{pmatrix}\;.
\eeq
We next color-decompose the Born amplitude for the process as 
\beq\label{eq:bornamplitude}
M^{(0)}=\sum_I h_I^{(0)}{\cal C}_I\;.
\eeq
Squaring the amplitude and summing  (averaging) over external colors and helicities, we find
\beq\label{H0S0new}
\frac{1}{4N_c^2}\sum_{a,b,c,d=1}^{N_c}|M^{(0)}|^2=\frac{1}{4N_c^2}
\sum_{a,b,c,d=1}^{N_c}\sum_{IJ}h_I^{(0)}h_J^{(0)*}{\cal C}^*_J{\cal C}_I=
\frac{1}{4N_c^2}\sum_{IJ}h_I^{(0)}h_J^{(0)*}S^{(0)}_{JI}\equiv{\mathrm{Tr}}[H^{(0)}S^{(0)}]\;,
\eeq
where
\beq\label{H0def}
(H^{(0)})_{IJ}\equiv \frac{1}{4N_c^2}\,h_I^{(0)}h_J^{(0)*}\;.
\eeq
While the matrix $H^{(0)}$ follows from a simple direct calculation, we extract it from
the results of~\cite{Kunszt:1993sd}, since we can then follow the same strategy for 
the one-loop results given there. The color-decomposed tree-level four-point helicity amplitudes 
for $qq'\rightarrow qq'$ are given in~\cite{Kunszt:1993sd} as
\ba\label{col1}
\mathcal{A}_{{\mathrm{tree}}}^{\lambda\lambda'} &=& 
\left(\delta_{ad}\,\delta_{bc}-\frac{1}{N_c}\,\delta_{ac}\,\delta_{bd}\right)\,
a_{4;0}^{\lambda\lambda'}\nn\\[2mm]
&\equiv& {\cal C}_{1} \times\left( 2 a_{4;0}^{\lambda\lambda'} \right) + {\cal C}_2 \times 0\;,
\ea
where $\lambda\lambda'$ denotes the helicity configuration of the initial partons. For a given
pair of helicity settings we have $h_{I=1}^{(0)}=2a_{4;0}$, $h_{I=2}^{(0)}=0$. The squares of the 
two helicity amplitudes are
\ba
|a_{4;0}^{--}|^2&=&\frac{s^2}{t^2}\;,\nn\\[2mm]
|a_{4;0}^{-+}|^2&=&\frac{u^2}{t^2}\;,
\ea
with the Mandelstam variables
\ba\label{stu_etanew}
s & = &(p_1+p_2)^2\,=\, \hat m^2 \;,\nn \\[2mm]
t & = &(p_1-p_3)^2\,=\, -\hat m^2 \f{e^{-\Delta\eta}}{e^{\Delta\eta}+e^{-\Delta\eta}}\;,\nn \\[2mm]
u & = & (p_1-p_4)^2\,=\,-\hat m^2 \f{e^{\Delta\eta}}{e^{\Delta\eta}+e^{-\Delta\eta}} \, .
\ea
Averaging over external colors and helicities appropriately, following Eq.~(\ref{H0S0new}), we obtain 
the lowest-order hard-scattering matrix as
\beq\label{H0}
H^{(0)}=
\begin{pmatrix}
\frac{2}{N_{c}^{2}}\frac{s^{2}+u^{2}}{t^2}& 0 \\[2mm] 0 & 0
\end{pmatrix}\,\equiv\,
\begin{pmatrix}
h_0& 0 \\[1mm] 0 & 0
\end{pmatrix}
\;,
\eeq
in agreement with~\cite{KO1}. As expected, its only entry is in the ``octet-octet'' corner, thanks to our
choice of color basis. 

\subsubsection{Hard part at one loop}

The hard-scattering matrix $H_{ab\rightarrow cd}$ is a perturbative function that contains all contributions 
associated with momenta of the order of the hard scale $\hat{m}$. Since in the threshold regime there is no 
phase space for hard on-shell radiation, only purely virtual contributions contribute to $H_{ab\rightarrow cd}$. 
Writing the virtual one-loop amplitude as (again we suppress the indices for the external particles) 
\beq\label{eq:1lamplitude}
M^{(1),{\mathrm{virt}}}=\sum_I \tilde h_I^{(1)}{\cal C}_I\;,
\eeq
and considering the interference with the Born amplitude, the elements of the first-order contribution 
$H^{(1)}_{ab\rightarrow cd}$ are obtained from the finite part of 
\beq\label{H1def}
(\tilde{H}^{(1)})_{IJ}\equiv\frac{1}{4N_c^2}\,\left( \tilde h_I^{(1)}h_J^{(0)*} + h_I^{(0)} \tilde h_J^{(1)*} \right)\;.
\eeq
Most of the one-loop amplitudes that we need are given in~\cite{Kunszt:1993sd}. 
For the gluonic channel $gg\rightarrow gg$, we additionally use the results 
of~\cite{Bern:1990cu,Bern:1991aq}. For the process $qq'\rightarrow qq'$, 
the one-loop four-point helicity amplitudes are given in~\cite{Kunszt:1993sd} as
\ba\label{col2}
\mathcal{A}_{{\mathrm{1loop}}}^{\lambda\lambda'} & = & 
\left(\delta_{ad}\,\delta_{bc}-\frac{1}{N_c}\,\delta_{ac}\,\delta_{bd}\right)\,
a_{4;1}^{\lambda\lambda'} +\delta_{ad}\,\delta_{bc} \, a_{4;2}^{\lambda\lambda'}\nn\\[2mm]
&=&{\cal C}_{1}\times 2\left(a_{4;1}^{\lambda\lambda'}+
a_{4;2}^{\lambda\lambda'}\right)+{\cal C}_2\times\frac{1}{N_c}a_{4;2}^{\lambda\lambda'}\;.
\ea
From this we can determine the $\tilde h_I^{(1)}$. Keeping in mind that we have pulled out
an overall factor $\as/\pi$ in our definition of the hard-scattering matrix $H^{(1)}$, cf. Eq.~(\ref{eq:Hexpansion}),
we have $\tilde h^{(1)}_{I=1}=2(a_{4;1}^{\lambda\lambda'}+a_{4;2}^{\lambda\lambda'})$ and 
$\tilde h^{(1)}_{I=2}= \,a_{4;2}^{\lambda\lambda'}/N_c$. As shown in~\cite{Kunszt:1993sd},
the $a_{4;1}^{\lambda\lambda'}$, $a_{4;2}^{\lambda\lambda'}$ are proportional to the tree-level 
$a_{4;0}^{\lambda\lambda'}$ in~(\ref{col1}) for each helicity configuration:
\ba
a_{4;1}^{\lambda\lambda'} &=& C_\Gamma\,
F^{\lambda\lambda'}_{4;1}a_{4;0}^{\lambda\lambda'}\;,\nn\\[2mm]
a_{4;2}^{\lambda\lambda'}  &=& C_\Gamma\, F^{\lambda\lambda'}_{4;2}a_{4;0}^{\lambda\lambda'}\;,
\ea
where in our normalization
\beq
C_\Gamma\;=\; \frac{e^{\gamma_E\varepsilon}}{4}\;\frac{\Gamma^2(1-\varepsilon)\;\Gamma(1+\varepsilon)}{\Gamma(1-2 \varepsilon)}\; .
\eeq
Here dimensional regularization with $D=4-2\varepsilon$ dimensions is used. The $F^{\lambda\lambda'}_{4;1}$, 
$F^{\lambda\lambda'}_{4;2}$ are functions of the Mandelstam variables. Using the shorthand notation
\beq\label{eq:logs1}
  L(t) = \log\frac{-t}{s}\,,  \;\; L(u) = \log\frac{-u}{s}\,,  \;\; L(s) = - i \pi  \,,
\eeq
we have from~\cite{Kunszt:1993sd}:
\ba
F_{4;1}^{--} & = & N_c\left[-\f{2}{\varepsilon^2}-\f{3}{\varepsilon}+2\f{L(s)}{\varepsilon}+
L^2(t)-\f{2}{3}L(t)\left(1+3L(s)\right)+\f{13}{9}+\pi^2 \right] +N_f\left[\f{2}{3}L(t)-\f{10}{9} \right] \nn\\
&& -\f{1}{N_c}\left[-\f{2}{\varepsilon^2}-\f{3}{\varepsilon}-\f{2}{\varepsilon}(L(s)-L(t)-L(u))-8-L^2(t) +
\f{u^2-s^2}{2s^2}\left((L(t)-L(u))^2+\pi^2\right)\right. \nn \\
&& \left.+\,2L(t)(1+L(s)-L(u))-\f{1}{s}(uL(t)+tL(u)) \right] +4\pi b_0\log\left(\f{\mu_R^2}{\hat m^2}\right)\, ,\nn \\
F_{4;1}^{-+} & = & N_c\left[-\f{2}{\varepsilon^2}-\f{3}{\varepsilon}+2\f{L(s)}{\varepsilon}+
L^2(t)-\f{2}{3}L(t)\left(1+3L(s)\right)+\f{13}{9}+\pi^2 \right] +N_f\left[\f{2}{3}L(t)-\f{10}{9} \right] \nn\\
&& -\f{1}{N_c} \left[-\f{2}{\varepsilon}-\f{3}{\varepsilon}-\f{2}{\varepsilon}(L(s)-L(t)-L(u)) -8-L^2(t)+L(t)(3+2L(s)-2L(u))\right] \nn \\
&&+\left(N_c+\f{1}{N_c}\right)\left[\f{s^2-u^2}{2u^2}(L^2(t)-2L(s)L(t))+\f{t}{u}(L(t)-L(s)) \right] +
4\pi b_0\log\left(\f{\mu_R^2}{\hat m^2}\right)\, ,\nn \\
F_{4;2}^{--} & = & 2C_F \left[\f{2}{\varepsilon} (L(u)-L(s))+\f{u^2-s^2}{2s^2}\left(L^2(t)+L^2(u)+\pi^2\right)+
\f{t}{s}(L(t)-L(u))+2L(s)L(t)\right.\nn \\
&& \left. \hspace{1.1cm} -\f{u^2+s^2}{s^2} L(t)L(u) \right] \, ,\nn \\
F_{4;2}^{-+} & = & 2C_F \left[\f{2}{\varepsilon} (L(u)-L(s))- \f{s^2-u^2}{2u^2}L^2(t)-
\f{t}{u}(L(t)-L(s))-2 L(t) L(u)+\f{s^2+u^2}{u^2} L(t)L(s) \right]\, . \nn \\ 
&&
\ea
Note that the loop corrections have imaginary parts arising from the analytic continuation of 
Mandelstam variables into the physical region $s>0;\;t,u<0$. They appear in the finite part as well 
as in the pole contributions.  

From this we can construct the matrix $\tilde H^{(1)}$ defined in Eq.~(\ref{H1def}) as
\begin{equation}
\tilde{H}^{(1)}= \frac{C_\Gamma}{4N_c^2}
\begin{pmatrix}
16\left(\mathcal{R}e\left(F^{--}_{4;2}+F^{--}_{4;1}\right)\frac{s^2}{t^2}+
\mathcal{R}e\left( F^{-+}_{4;2}+F^{-+}_{4;1}\right)\frac{u^2}{t^2}\right) &\frac{4}{N_c}
\left({F^{--}_{4;2}}^* \frac{s^2}{t^2}+{F^{-+}_{4;2}}^* \frac{u^2}{t^2}\right)  \\[3mm]
\frac{4}{N_c}\left({F^{--}_{4;2}} \frac{s^2}{t^2}+{F^{-+}_{4;2}} \frac{u^2}{t^2}\right) &0
\end{pmatrix}\;.
\end{equation}
The full expression for this matrix is rather lengthy. It has the following explicit structure:
\ba\label{MV}
\tilde{H}^{(1)}&=&\frac{1}{2}\,\left[
\left(-\frac{4 C_F}{\varepsilon^2}-\frac{6 C_F}{\varepsilon}\right)\,H^{(0)}
-\frac{L(s)}{\varepsilon}
\frac{C_F}{N_c}\,
\begin{pmatrix}0 & -1 \\[2mm] 1 & 0
\end{pmatrix}-\frac{2}{\varepsilon}L(t)\frac{h_0}{N_c}
\begin{pmatrix}1 & 0 \\[2mm] 0 & 0
\end{pmatrix}\right.\nn\\[2mm]
&+&\frac{1}{\varepsilon}L(u)\frac{h_0}{N_c}
\begin{pmatrix}2 (N_c^2-2) & C_F \\[2mm] C_F & 0
\end{pmatrix}+
%\frac{1}{\varepsilon}\log\left(\frac{s}{s}\right)\frac{h_0}{N_c}
%\begin{pmatrix}4 & -C_F \\[2mm] -C_F & 0
%\end{pmatrix} +
4\pi b_0 \log\left(\frac{\mu_R^2}{\hat{m}^2}\right)H^{(0)} \Bigg] 
%\nn \\[2mm]
%&&
+  \,H^{(1)}\;,
\ea
where $h_0$ and $H^{(0)}$ have been given in~(\ref{H0}). 
Following~\cite{Sterman:2002qn}, we have identified the finite part in the last line with the 
first-order correction to the hard-scattering matrix. This finite part is a function of the Mandelstam 
variables only. As one can see, the explicit dependence on the renormalization scale $\mu_R$ has been 
separated from $H^{(1)}$. It is proportional to $H^{(0)}$ and therefore fully taken into account 
by the exponential $\xi_R$ in Eq.~(\ref{xi11}), as discussed in Sec.~\ref{sec32}.

To present our final results for $H^{(1)}$, we adopt the notation of Ref.~\cite{Kelley:2010fn},
where the matrix was derived in the context of the soft-collinear effective theory. 
We find full agreement with the result in their Eq.~(39):
\ba\label{H1final}
\left(H^{(1)}\right)_{11} &=&  \;\mathcal{R}e\;\Bigg\{ \frac{1}{2 N_c^2}
\Big[\frac{s^2+u^2}{t^2} \left( -4 C_F L(t)^2 + 2 X_1 (s,t,u) L(t)+2 Y\right)
\nn \\[2mm]
&&\hspace*{1.2cm}+\frac{s^2}{t^2}(C_A-4C_F) Z(s,t,u) -\frac{u^2}{t^2} (2C_A-4C_F) Z(u,t,s)\Big]\Bigg\}\;,\nn\\[2mm]
\left(H^{(1)}\right)_{21} &=& \frac{1}{2 N_c^2}
 \Big[\frac{s^2+u^2}{t^2} X_2 (s,t,u) L(t)
-\frac{s^2}{t^2}\frac{C_F}{2C_A}Z(s,t,u) +\frac{u^2}{t^2}\frac{C_F}{2C_A} Z(u,t,s)\Big] \;,\nn\\[2mm]
\left(H^{(1)}\right)_{12} &=& \left(H^{(1)}\right)^*_{21} \;,\nn \\[2mm]
\left(H^{(1)}\right)_{22} &=& 0\, ,\label{H1fin}
\ea
with~\cite{Kelley:2010fn}
\ba
X_1(s,t,u) &= &
\phantom{\frac{C_F}{C_A}\!\!\!\!\!\!\!\!} 6 C_F - 4 \pi  b_0 + 8 C_F [L(s) - L(u)] - 2C_A[2L
(s)-L(t)-L(u)]\,,\nn \\[2mm]
X_2(s,t,u) &=&
\frac{2C_F}{C_A} [L(s)-L(u)]\,,\nn\\[2mm]
Y &=&C_A\left(\frac{10}{3} + \pi^2\right) + C_F\left(\frac{\pi^2}{3} - 16\right) + \frac{5}{3} 4 \pi  b_0 \,,\nn\\[2mm]
Z(s,t,u) &= &\frac{t}{s}\left(\frac{t+2 u}{s}[L(u) - L(t)]^2 +2[ L(u)-L(t)] + \pi^2\frac{t
 + 2 u }{s}\right)\,.
\ea
There are several ways of checking the validity of the results. The simplest one is to 
compute 
\be
{\mathrm{Tr}}\left[\tilde{H}^{(1)}\, S^{(0)}\right] \;, 
\ee
which 
should reproduce the known one-loop virtual correction to $qq'\to qq'$ scattering given in~\cite{Kunszt:1993sd}. 
This indeed turns out to be the case. Since $S^{(0)}$ is diagonal in our basis, this provides a check on 
the diagonal elements of $H^{(1)}$. 

We also note that the pole terms of the NLO virtual amplitudes $M^{(1),{\mathrm{virt}}}$ in~(\ref{eq:1lamplitude}), 
including their imaginary parts, have been predicted in~\cite{Sterman:2002qn,Catani:1998bh} to be given by 
\be\label{StYeo}
\tilde h_I^{(1)}|_{\begin{subarray}{l} {\mathrm{pole}} \\ {\mathrm{terms}} \end{subarray}}
= \f{1}{2}\left[-C_F\left(\frac{2}{\varepsilon^2}+\frac{3}{\varepsilon}\right)\mathbb{1}+\frac{1}{\varepsilon}
\Gamma_{qq'\to qq'}^{(1)} \right]_{IJ}h_J^{(0)}\, ,
\ee
where $\mathbb{1}$ denotes the $2\times 2$ unit matrix and 
$\Gamma^{(1)}_{qq'\to qq'}$ is the soft anomalous dimension matrix introduced in~(\ref{GammaSoft})
which possesses imaginary parts (the explicit result for $\Gamma^{(1)}_{qq'\to qq'}$ is 
given in Eq.~(\ref{eq:Gammaexplicit}) below). We have verified that this correctly reproduces the pole terms in the 
$\tilde h_I^{(1)}$. 

In the way described in this subsection we have determined the one-loop hard-scattering matrix for each 
partonic channel contributing to di-hadron production. As one can see in Eqs.~(\ref{MV}),(\ref{H1fin}), 
for $qq'\to qq'$ the final expression always contains the squares of the tree-level helicity amplitudes $a_{4;0}$. 
This becomes different for partonic channels involving both external quarks and gluons. 

\subsection{Soft function \label{SF}}

We now turn to the computation of the first-order correction $S_{ab\to cd}^{(1)}$ 
to the soft function in Eq.~(\ref{eq:soft1}). Again we present explicit results only
for the $qq'$ channel, although we have of course considered all partonic channels.
In fact, in the course of the study of $qq'$ scattering we find a general construction 
rule for the soft matrix $S^{(1)}_{ab\to cd}$ that turns out to be applicable to all partonic 
channels. 

\subsubsection{Color structure of diagrams in the eikonal approximation}

\begin{figure}[t]
\centering
\vspace*{-1.2cm}
%\hspace*{-6mm}
\epsfig{figure=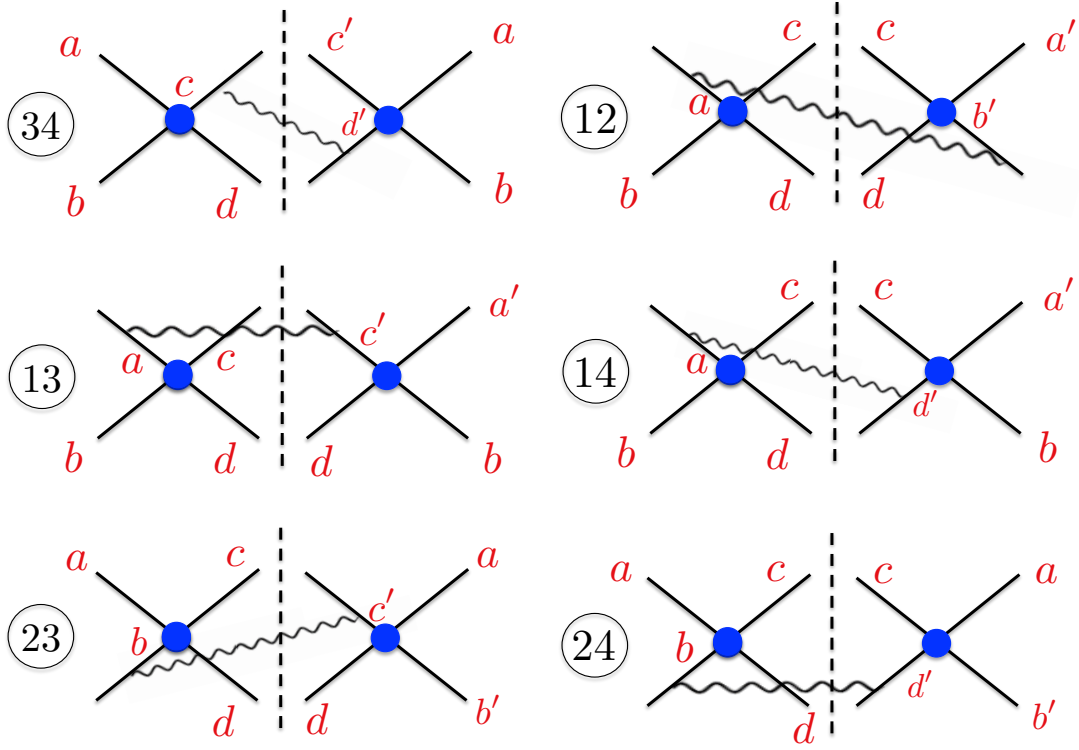,width=0.9\textwidth}

\vspace*{-1.4cm}
\caption{\sf Diagrams relevant for the calculation of the NLO soft matrix $S^{(1)}$. 
The blobs represent a Born amplitude and the letters are color indices.\label{fig:S1}}
\vspace*{0.cm}
\end{figure}

In order to to compute the soft matrix at NLO for $qq'$ scattering, we need to consider the
process $q(p_1)+q'(p_2)\rightarrow q(p_3)+q'(p_4)+g(k)$, where $g$ denotes a radiated gluon 
with soft momentum $k$. The diagrams are treated in the eikonal approximation, decomposed
according to our color basis. They are shown in Fig.~\ref{fig:S1}. The blobs on either side of 
the cut denote a Born hard part that can be a color-octet or a singlet. There are six diagrams 
labeled ``34'' or ``12'' for example, depending on the external legs between which the additional gluon is 
exchanged. Eventually, all contributions must be summed. Using the notation of the previous subsection, 
each of the diagrams in Fig.~\ref{fig:S1} has the structure
\beq
\sum_{IJ}h_I^{(0)}h_J^{(0)*}\,({\cal R}_{ij})_{JI}\,I_{ij}\;,
\eeq
where $ij$ labels the diagram, $I_{ij}$ is an integral over the eikonal factor corresponding to the diagram that
we will specify below, and the $({\cal R}_{ij})_{JI}$ form a $2\times 2$ matrix with entries labeled by $JI=$octet-octet, 
singlet-octet, etc. For example, for the ``octet-octet'' entry of ${\cal R}_{34}$ we have
\ba
({\cal R}_{34})_{\begin{subarray}{c} J={\mathrm{octet}} \\
I={\mathrm{octet}}\end{subarray}}
&=&\sum_{\begin{subarray}{c}
       a,b,c,d,c',d' \\[0mm] g,g_1,g_2
      \end{subarray}}T^{g_2}_{bd'}T^{g_2}_{ac'}\;T^g_{d'd}T^g_{c'c}\;T^{g_1}_{db}T^{g_1}_{ca}\nn\\[2mm]
      &=&\sum_{g,g_1,g_2}{\mathrm{Tr}}[T^{g_2}T^gT^{g_1}]\,{\mathrm{Tr}}[T^{g_2}T^gT^{g_1}]\nn\\[2mm]
      &=&-\frac{N_c^2-1}{4N_c}=-\frac{C_F}{2}\;.
\ea
Here $g$ corresponds to the color of the gluon exchanged between the external legs, while $g_1$ and $g_2$ 
are those for the gluons in the amplitudes on the two sides of the cut. Computing in this way the matrices 
${\cal R}_{ij}$ for all diagrams, we find:
\ba
{\cal R}_{12}&=&{\cal R}_{34}\,=\,\frac{C_F}{2}
\left(
   \begin{array}{cc}
   -1& N_c \\[2mm]
 N_c & 0
   \end{array}
   \right)\;,\nn\\[2mm]
   {\cal R}_{13}&=&{\cal R}_{24}\,=\,\frac{C_F}{2}
\left(
   \begin{array}{cc}
   -\frac{1}{2}& 0 \\[2mm]
0 & 2 N_c^2
   \end{array}
   \right)\;,\nn\\[2mm]
     {\cal R}_{14}&=&{\cal R}_{23}\,=\,\frac{C_F}{2}
\left(
   \begin{array}{cc}
   \frac{1}{2}(N_c^2-2)& N_c \\[2mm]
N_c & 0
   \end{array}
   \right)\;.
\ea
For the sum of all diagrams we thus have
\beq\label{R1}
%\hspace*{-.3cm}
\sum_{ij}\,{\cal R}_{ij}\,I_{ij}\,=\,\frac{C_F}{2}
\left(
   \begin{array}{cc}
   \frac{1}{2}(I_{13}+I_{24})-I_{12}-I_{34}-\frac{N_c^2-2}{2}(I_{14}+I_{23})& N_c(I_{12}+I_{34}-
   I_{14}-I_{23}) \\[2mm]
N_c(I_{12}+I_{34}-
   I_{14}-I_{23}) &
   -2 N_c^2 (I_{13}+I_{24})
   \end{array}
   \right)\, .
\eeq
We note that the eikonal factor for the interference between initial- and final-state emission has an extra minus 
sign which we included here.
%%%%%%%%%%%%%%%%%%%%%%%%%%%%%%%%%%%%%%%%%%%%%%%%%%%%%%%%
%%%%%%%%%%%%%%%%%%%%%%%%%%%%%%%%%%%%%%%%%%%%%%%%%%%%%%%%

\subsubsection{Integrals $I_{ij}$}

Next, we need to specify and compute the $I_{ij}$. They are essentially given by eikonal factors integrated 
over the gluon phase space. They are normalized relative to the Born cross section. Adopting the 
three-particle phase space in $D=4-2\varepsilon$ dimensions from~\cite{Almeida:2009jt} 
(see also~\cite{Czakon:2009zw}), one has
\begin{equation}
I_{ij} = -\f{\as}{\pi}\f{s}{4\pi} e^{\varepsilon \gamma_E}\frac{\Gamma(1-\varepsilon)}{\Gamma(1-2\varepsilon)}
\int_0^1 d\hat{\tau}\,\hat{\tau}^{-\varepsilon}(1-\hat{\tau})^{1-2\varepsilon}
\int d\Omega\, {p_i\cdot p_j \over (p_i\cdot k)(p_j\cdot k)}\, ,
\end{equation}
where
\ba
\int d\Omega=\int_0^{\pi} d\psi \sin^{1-2\varepsilon}\psi \int_0^\pi d\theta\,
\sin^{-2 \varepsilon}\theta \; .
\ea
We work in the c.m.s. of the incoming partons; $\psi$ and $\theta$
are the gluon's polar and azimuthal angles relative to the plane
defined by the directions of incoming and outgoing hard partons.
The relevant angular integrals are well-known~\cite{WvN}:
\ba
\label{omega}
&&\int d\Omega \,
\frac{1}{(1-\cos\psi)^j (1-\cos\psi \cos\chi-\sin\psi \cos\theta \sin\chi)^k} \nn \\[2mm]
&&\hspace{1cm}= 2 \pi \frac{\Gamma(1-2 \varepsilon)}{\Gamma(1-\varepsilon)^2}
\,2^{-j-k}\,B(1-\varepsilon-j,1-\varepsilon-k)\, {}_2 F_1 \left(j,k,1-\varepsilon,\cos^2 \frac{\chi}{2} \right) \;,
\ea
with the Hypergeometric function ${}_2 F_1$. Performing the integrations over $d\Omega$, 
but leaving the integration over $\hat\tau$ (or, equivalently, gluon energy) aside for the moment, 
we find near threshold
\ba
&&\frac{dI_{12}}{d\hat\tau}=\frac{dI_{34}}{d\hat\tau}=-\frac{\alpha_s}{\pi}
\left[ \left(\frac{1}{\varepsilon^2}-\frac{\pi^2}{4}\right)\delta(1-\hat\tau)-\frac{2}{\varepsilon}
\,\frac{1}{(1-\hat\tau)_+}+4\left(\frac{\ln(1-\hat\tau)}{1-\hat\tau}\right)_+\right]\,,\nn\\[2mm]
&&\frac{dI_{13}}{d\hat\tau}=\frac{dI_{24}}{d\hat\tau}=-\frac{\alpha_s}{\pi}
\left[ \left(\frac{1}{\varepsilon^2}-\frac{\pi^2}{4}-\frac{1}{\varepsilon}\ln\left(-\frac{t}{s}\right)-
{\mathrm{Li}}_2\left(-\frac{u}{t}\right)
\right)\delta(1-\hat\tau)\right.\nn\\[2mm]
&&\hspace*{3.2cm}\left.
+2\left(-\frac{1}{\varepsilon}+\ln\left(-\frac{t}{s}\right)\right)
\,\frac{1}{(1-\hat\tau)_+}+4\left(\frac{\ln(1-\hat\tau)}{1-\hat\tau}\right)_+\right]\,,\nn\\[2mm]
&&\frac{dI_{23}}{d\hat\tau}=\frac{dI_{14}}{d\hat\tau}=\frac{dI_{13}}
{d\hat\tau}\Big|_{t\leftrightarrow u}\,,\label{eq:ls}
\ea
where ${\mathrm{Li}}_2$ denotes the Dilogarithm function.
 
\subsubsection{Extraction of $S^{(1)}$}

Combining Eqs.~(\ref{R1}) and~(\ref{eq:ls}), we obtain
\ba\label{R2}
&&\hspace*{-1.4cm}\sum_{ij}\,{\cal R}_{ij}\,\frac{dI_{ij}}{d\hat\tau}\nn\\[2mm]
&=&C_F\frac{\alpha_s}{\pi}\Bigg[\left\{\delta(1-\hat\tau)\,\left(\frac{1}{2\varepsilon^2}-\frac{\pi^2}{8}\right)-\frac{1}{\varepsilon}
\,\frac{1}{(1-\hat\tau)}_++2\left(\frac{\ln(1-\hat\tau)}{1-\hat\tau}\right)_+
\right\}\left(
   \begin{array}{cc}
   N_c^2-1 & 0 \\[2mm]
0 &4 N_c^2
   \end{array}
   \right)\nn\\[2mm]   
&&\hspace*{9mm}+\,\left\{\delta(1-\hat\tau)\,\frac{1}{2\varepsilon}-\frac{1}{(1-\hat\tau)}_+\right\}
\left(
   \begin{array}{cc}
   \ln\left(-\frac{t}{s}\right) +(2-N_c^2)\ln\left(-\frac{u}{s}\right)& -2N_c\ln\left(-\frac{u}{s}
   \right) \\[2mm]
-2N_c\ln\left(-\frac{u}{s}\right) & -4 N_c^2\ln\left(-\frac{t}{s}\right)
   \end{array}
   \right)\nn\\[2mm]
&&\hspace*{9mm}+\,\delta(1-\hat\tau)\,\frac{1}{2}
    \left(
   \begin{array}{cc}
   {\mathrm{Li}}_2\left(-\frac{u}{t}\right)+
   (2-N_c^2)\,{\mathrm{Li}}_2\left(-\frac{t}{u}\right)&
   -2 N_c\, {\mathrm{Li}}_2\left(-\frac{t}{u}\right)
   \\[2mm]
-2 N_c\, {\mathrm{Li}}_2\left(-\frac{t}{u}\right) & -4 N_c^2\,
 {\mathrm{Li}}_2\left(-\frac{u}{t}\right)
   \end{array}
   \right)\Bigg]\;.
\ea
In the first term we recognize the lowest-order soft matrix of Eq.~(\ref{S0mat}). 
The matrix in the second term has a direct relation to the one-loop soft anomalous
dimension matrix $\Gamma_{qq'\to qq'}^{(1)}$ introduced 
in~(\ref{GammaSoft}),(\ref{eq:Gamma12}), which in our color basis is given by~\cite{KO1}
\be\label{eq:Gammaexplicit}
\Gamma^{(1)}_{qq'\to qq'}=
\begin{pmatrix}
\frac{1}{N_{c}}(2S-T-U)+2C_F U & 2 (U-S) \\[2mm]
\frac{C_{F}}{N_{c}}(U-S) & 2C_F T
\end{pmatrix}\,,
\ee
with\footnote{An equally possible choice is 
$S= - i\pi,T=\ln\left(\frac{-t}{s}\right), U=\ln\left(\frac{-u}{s}\right)$, which matches the logarithms 
in Eq.~(\ref{eq:logs1}). One easily checks that $\Gamma_{qq' \to qq'}^{(1)}$ differs for the two 
choices only by a term proportional to the unit matrix which commutes with all other matrices and 
hence cancels in the final result. This holds true for all partonic channels.
Note that even for the resummed cross section
only the combination ${\mathrm{e}}^{\Gamma^\dagger}S{\mathrm{e}}^{\Gamma}$ contributes.}
\be\label{eq:STU2}
S= 0, \quad T=\ln\left(\frac{-t}{s}\right)+i\pi,\quad U=\ln\left(\frac{-u}{s}\right)+i\pi \;.
\ee
One easily checks that the matrix in the second line of the right-hand-side 
of~(\ref{R2}) is given by $-[(\Gamma^{(1)})^{\dagger}S^{(0)}+S^{(0)}
\Gamma^{(1)}]/C_F$. Hence, we have after some reordering of terms:
\ba\label{R3}
\sum_{ij}\,{\cal R}_{ij}\,\frac{dI_{ij}}{d\hat\tau}&=&\frac{\alpha_s}{\pi}\Bigg[\delta(1-\hat\tau)\,
\left\{\frac{2C_F}{\varepsilon^2}S^{(0)}-\frac{1}{2\varepsilon}\,
\left[(\Gamma^{(1)})^{\dagger}S^{(0)}+S^{(0)}\Gamma^{(1)}\right]\right\}
\nn\\[2mm]
&&\hspace*{-1cm}-\,\frac{4C_F}{\varepsilon}\,\frac{1}{(1-\hat\tau)}_+S^{(0)}
+8C_F\left(\frac{\ln(1-\hat\tau)}{1-\hat\tau}\right)_+S^{(0)}
+\frac{1}{(1-\hat\tau)}_+\left[(\Gamma^{(1)})^{\dagger}S^{(0)}+S^{(0)}
\Gamma^{(1)}\right]\nn\\[2mm]
&&\hspace*{-1cm}+\,\frac{C_F}{2}\delta(1-\hat\tau)\left\{\left(
   \begin{array}{cc}
   {\mathrm{Li}}_2\left(-\frac{u}{t}\right)+
   (2-N_c^2)\,{\mathrm{Li}}_2\left(-\frac{t}{u}\right)&
   -2 N_c\, {\mathrm{Li}}_2\left(-\frac{t}{u}\right)
   \\[2mm]
-2 N_c\, {\mathrm{Li}}_2\left(-\frac{t}{u}\right) & -4 N_c^2\,
 {\mathrm{Li}}_2\left(-\frac{u}{t}\right)
   \end{array}
   \right)-\pi^2 S^{(0)}\right\}\Bigg] \;.\nn\\[2mm]
\ea
Each of the terms in this equation has a transparent interpretation. The pole terms $\propto \delta
(1-\hat\tau)$ in the first line will be canceled by corresponding terms in the virtual correction;
see Eq.~(\ref{StYeo}). The single pole term $\propto 1/(1-\hat\tau)_+$ will be canceled 
by collinear factorization in the eikonal approximation, as described in the~Appendix. 
The next two terms precisely match
the threshold logarithms at NLO, as becomes evident by going to Mellin-moment space and
comparing to~(\ref{nloexp}). The remaining term involves the one-loop soft matrix we are
interested in. More precisely, since $S^{(1)}$ appears in the Mellin-space expression 
for the resummed cross section, and since the moments of $(\ln(1-\hat\tau)/(1-\hat\tau))_+$ are 
given by $\frac{1}{2}(\ln^2\bar N+\pi^2/6)$ (up to corrections suppressed as $1/N$), all
terms $\propto \pi^2$ match when comparing to~(\ref{nloexp}), and we are just left with
\beq
S^{(1)}\,=\,\frac{C_F}{2}\left(
   \begin{array}{cc}
   {\mathrm{Li}}_2\left(-\frac{u}{t}\right)+
   (2-N_c^2)\,{\mathrm{Li}}_2\left(-\frac{t}{u}\right)&
   -2 N_c\, {\mathrm{Li}}_2\left(-\frac{t}{u}\right)
   \\[2mm]
-2 N_c\, {\mathrm{Li}}_2\left(-\frac{t}{u}\right) &-4 N_c^2\,
 {\mathrm{Li}}_2\left(-\frac{u}{t}\right)
   \end{array}
   \right)\;.
\eeq
This is our final result for the one-loop soft matrix for this process. A powerful check on the
result comes from comparison with the full cross section at NLO: Inserting our $S^{(1)}$ along
with $H^{(1)}$ from Eq.~(\ref{H1final}) into~(\ref{nloexp}), we verify that the
resulting expression correctly  reproduces all threshold logarithms {\it and all constant terms} 
in the NLO partonic cross section. 

As it turns out, we can give a very simple rule for obtaining $S^{(1)}$ directly from $S^{(0)}$ and  the anomalous dimension
matrix $\Gamma^{(1)}$. This becomes already evident from comparison of the two matrices in the second and third lines 
of~(\ref{R2}): They have identical structure, except that each logarithm has to be replaced by a dilogarithm with
suitably modified argument,
\ba\label{logmod}
\ln\left(-\f{t}{s}\right) & \rightarrow & \mathrm{Li}_2\left(-\f{u}{t} \right) \;, \nn \\[2mm]
\ln\left(-\f{u}{s} \right) & \rightarrow & \mathrm{Li}_2\left(-\f{t}{u} \right) \; .
\ea
The deeper reason for this is of course that already in the integrals~(\ref{eq:ls}) the logarithm
and the dilogarithm always appear in the same ratio in the term $\propto\delta(1-\hat\tau)$. 
Since we know how the matrix in the second line of~(\ref{R2}) is expressed in terms of $S^{(0)}$ and 
$\Gamma^{(1)}$, we also know how to construct $S^{(1)}$: Compute the combination  
$-1/2\left((\Gamma^{(1)})^{\dagger}S^{(0)}+S^{(0)}\Gamma^{(1)}\right)$ and 
substitute each logarithm according to~(\ref{logmod}). 
As the integrals $I_{ij}$ are the same no matter which process we are considering, 
this simple construction rule works for all partonic channels. All necessary ingredients,
the $\Gamma^{(1)}_{ab\to cd}$ and the $S^{(0)}_{ab\to cd}$, may be found in the Appendix of Ref.~\cite{KO1};
we therefore do not present the explicit expressions for the resulting $S^{(1)}_{ab\to cd}$ for all the other
channels, which become rather lengthy. It is likely that the simple rule we find is a special 
property of the pair mass kinematics we are considering here.

\subsection{Inverse Mellin and Fourier transforms and matching procedure}

In order to produce phenomenological results for the resummed case, we need 
to perform inverse Mellin transform and Fourier transforms. The Mellin inverse 
requires a prescription for dealing with the singularity in the perturbative strong 
coupling constant in the NNLL expansions of the resummed exponents. As
in~\cite{Almeida:2009jt} we will use the {\em Minimal Prescription} developed in~\cite{Catani:1996yz}, 
which relies on use of the NNLL expanded forms 
given in Sec.~\ref{sec32} and on choosing a Mellin contour in complex-$N$ space that lies to the {\it left}
of the poles at $\lambda=1/2$ and $\lambda=1$ in the Mellin integrand.
The function $\Omega_{H_1 H_2\to cd}$ in~(\ref{Dconv}) is obtained as~\cite{Almeida:2009jt}
\beqa \label{Omegainv}
&&\hspace{-1.2cm}\Omega_{H_1 H_2\to cd}^{\mathrm{resum}}
\left( \tau', \Delta \eta, \bar{\eta},  \as(\mu_R), \frac{\mu_R}{\hat{m}}, \frac{\mu_F}{\hat{m}}
\right) =\frac{1}{2\pi}\int_{-\infty}^{\infty} 
d\nu \, {\mathrm{e}}^{-i \nu \bar{\eta}}
\int_{C_{MP}-i\infty}^{C_{MP}+i\infty} \frac{dN}{2\pi i}\,
\left(\tau'\right)^{-N} \nn \\[2mm]
&&\hspace{-0.6cm}\times \sum_{ab} 
\tilde{f}_a^{H_1}(N+1+i\nu/2,\mu_F)\tilde{f}_b^{H_2}(N+1-i\nu/2,\mu_F)
\; \tilde{\omega}_{ab\to cd}^{\mathrm{resum}}
\left(N,\nu, \Delta \eta, \as(\mu_R),\frac{\mu_R}{\hat{m}}, \frac{\mu_F}{\hat{m}} \right)  , 
\eeqa
with a suitable Mellin contour consistent with the minmal prescription. 
As shown in~\cite{Almeida:2009jt}, it is straightforward to perform the 
convolution of the inverted resummed $\Omega_{H_1 H_2\to cd}^{\mathrm{resum}}$ 
with the fragmentation functions, as given by~(\ref{Omegainsigma}). 

As in~\cite{Almeida:2009jt}, we match the resummed cross section 
to the full NLO one, by expanding the resummed 
cross section to ${\cal O}(\as^3)$, subtracting the expanded result
from the resummed one, and adding the full NLO cross section:
\beq
\label{hadnres}
d \sigma^{{\mathrm{match}}} = \left( d \sigma^{{\mathrm{resum}}}-
d \sigma^{{\mathrm{resum}}}\Big|_{{\cal O}(\alpha_s^3)}\right)+
d\sigma^{\mathrm{NLO}} \; .
\eeq
For the NLO cross section we use the results of~\cite{Owens:2001rr}. 
In this way, NLO is taken into account in full, and the soft-gluon 
contributions beyond NLO are resummed in the way described
above. Of course, for a full NNLL resummed cross section one would prefer
to match to an NNLO calculation, which however is not available for this
observable yet.

\section{Phenomenological results \label{sec4}}

We now examine the numerical effects of our approximate NNLL resummation in comparison to the
NLL and NLO results shown in~\cite{Almeida:2009jt}. Since the NNLL effects are generally rather
similar for the experimental situations considered in~\cite{Almeida:2009jt}, we show only two
representative examples here. We will also make predictions for the di-hadron cross section at 
RHIC, where one would expect the effects of resummation to be smaller.

Our examples from~\cite{Almeida:2009jt} concern the NA24~\cite{na24} and the CCOR~\cite{ccor}
$pp\to \pi^0\pi^0$ scattering experiments. The fixed-target experiment NA24 recorded data 
at a beam energy of $E_p=300$~GeV, while CCOR operated at the ISR collider at $\sqrt{S}=62.4$~GeV.
Both experiments employed the cuts $p_T^{\mathrm{pair}}<1$~GeV, $|Y|<0.35$, and $|\cos\theta^*|<0.4$. 
Here,  $p_T^{\mathrm{pair}}$ and $Y$ are the transverse momentum and rapidity of the pion pair, respectively, 
which are given in terms of the individual pion transverse momenta $p_{T,i}$ and of $\Delta\eta,\bar\eta$ in~(\ref{baretadef1}) 
by
\beqa
p_T^{\mathrm{pair}}&=&|p_{T,1}-p_{T,2}| \; ,\nn\\[2mm]
Y&=&\bar{\eta} - \frac{1}{2} \ln\left( \frac{p_{T,1}\,{\mathrm{e}}^{-\Delta\eta} +
p_{T,2}\,{\mathrm{e}}^{\Delta\eta}}{p_{T,1}\,{\mathrm{e}}^{\Delta\eta} +
p_{T,2}\,{\mathrm{e}}^{-\Delta\eta}} \right) \; ,
\eeqa
where  LO kinematics have been assumed as appropriate in the threshold regime.
Furthermore, $\cos\theta^*$ is the cosine of the scattering angle in the partonic c.m.s.
and is for LO kinematics given by
\begin{equation}
\cos\theta^*=\frac{1}{2}
\left( \frac{p_{T,1}}{p_{T,2}+p_{T,1}\cosh(2 \Delta\eta)} +
\frac{p_{T,2}}{p_{T,1}+p_{T,2}\cosh(2 \Delta\eta)} \right) 
\sinh(2\Delta\eta) \; . \label{ctdef}
\end{equation}
For details on the kinematical variables, see~\cite{Almeida:2009jt}.
Thanks to our way of organizing the threshold resummed cross 
section, inclusion of cuts on any of these variables is
straightforward.

In all our calculations, we use the CTEQ6M5 set of parton distribution functions~\cite{cteq6}, 
along with its associated value of the strong coupling constant. As compared to our
results in~\cite{Almeida:2009jt}, we update to the latest ``de Florian-Sassot-Stratmann'' (DSS) set of
fragmentation functions~\cite{DSS}. We note that one might object that the use of NLO parton 
distribution functions and fragmentation functions is not completely justified for obtaining NNLL 
resummed predictions. However, since fragmentation functions evolved at NNLO
are not yet available in any case, we have decided to stick to NLO functions throughout. As in~\cite{Almeida:2009jt}, 
we choose the renormalization and factorization scales to be equal, $\mu_R=\mu_F\equiv\mu$, and 
we give them the values $M$ and $2M$, in order to investigate the scale dependence of the results. 
%%%%%%%%%%%%%%%%%%%%%%%%%%%%%%%%%%%%%%%%%%%%%%%%%%%%%%%%
\begin{figure}[t]
\begin{center}
\vspace*{-1.8cm}
\hspace*{-1.6cm}
\epsfig{figure=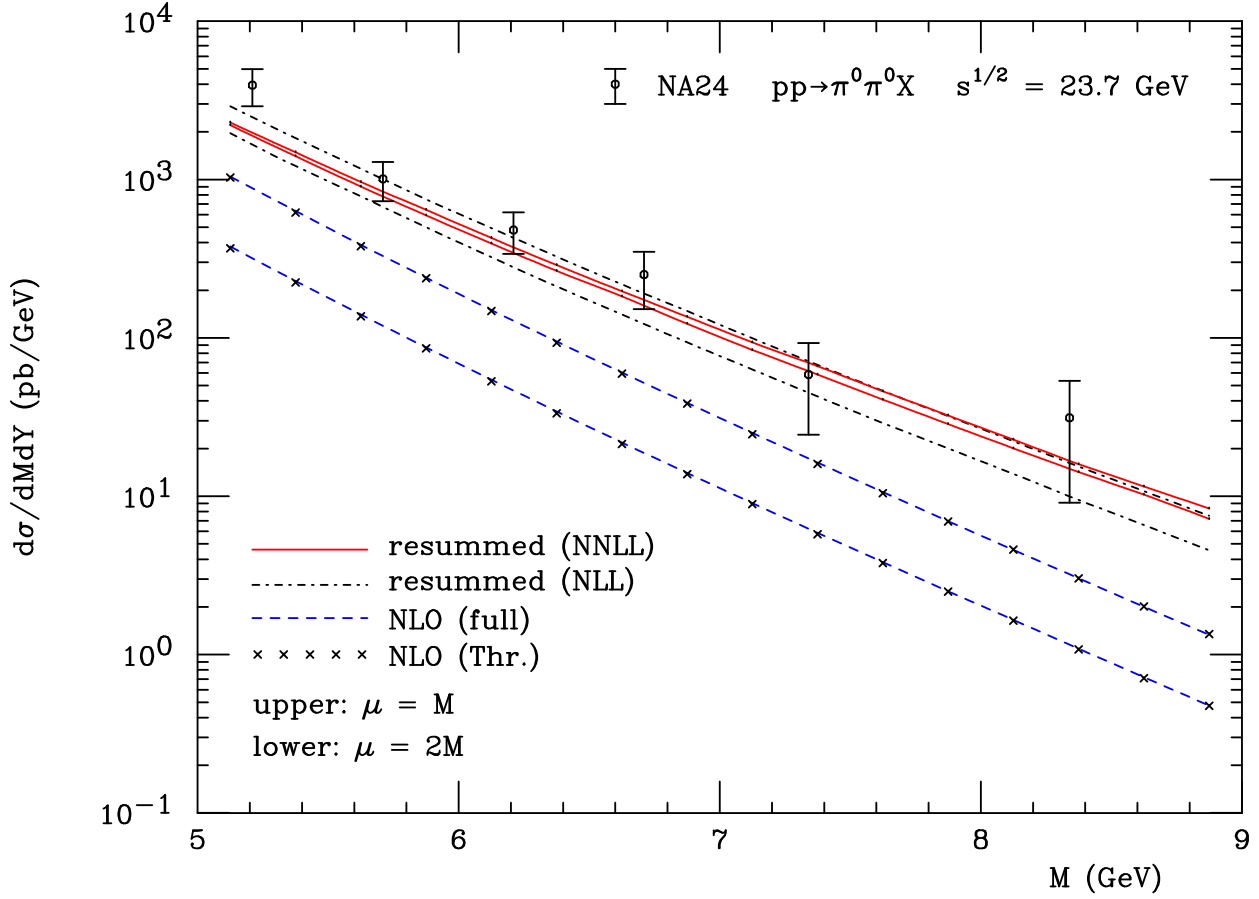,width=0.9\textwidth,angle=90}
\end{center}
\vspace*{-2cm}
\caption{\sf  Comparison of NLO (dashed), NLL resummed (dot-dashed) and NNLL resummed (solid) 
calculations of the cross section for $pp\to \pi^0\pi^0X$ to the NA24 data~\cite{na24}, for two different choices of 
the renormalization and factorization scales, $\mu_R=\mu_F=M$ (upper lines) and $\mu_R=\mu_F=2M$ (lower lines). 
The crosses display the NLO ${\cal O}(\alpha_s)$ expansion of the resummed cross section. 
\label{fig:na24nnll}}
\vspace*{0.1cm}
\end{figure}
%%%%%%%%%%%%%%%%%%%%%%%%%%%%%%%%%%%%%%%%%%%%%%%%%%%%%%%%

Figure~\ref{fig:na24nnll} shows the comparison to the NA24~\cite{na24} data. As known from~\cite{Almeida:2009jt},
the full NLO cross section and the first-order expansion of the resummed expression, that is, the last two terms in 
Eq.~(\ref{hadnres}), agree to a remarkable degree. Their difference actually never exceeds $1\%$ for the kinematics relevant for NA24. 
We recall these results by the dashed lines and the crosses in the figure. They provide confidence that the soft-gluon terms constitute 
the dominant part of the cross section, so that their resummation is sensible. The dot-dashed lines in the figure present the 
NLL results, computed by dropping all NNLL terms and matching to NLO via Eqs.~(\ref{Ccoeff1}),(\ref{Ccoeff}),
as in~\cite{Almeida:2009jt}. As found there, resummation leads to a significant enhancement of the 
theoretical prediction and provides a much better description of the NA24 data~\cite{na24} than for the NLO calculation.
Finally, the two solid lines show our NNLL resummed results. The key observations are that the two NNLL results
for scales $2M$ and $M$ are very close together and both roughly fall within the ``band'' spanned by the two NLL results 
for the two scales. One also notices that the NNLL curves have a slope somewhat less steep than the NLL ones. 
Given the relatively large uncertainties of the data, it is fair to say that the main effects are already taken into account at 
NLL. However, the precision of the NNLL calculation, in particular the strong reduction of the scale dependence, still
provides a significant theoretical and phenomenological improvement.  

\begin{figure}[t]
\begin{center}
\vspace*{-1.8cm}
\hspace*{-1.6cm}
\epsfig{figure=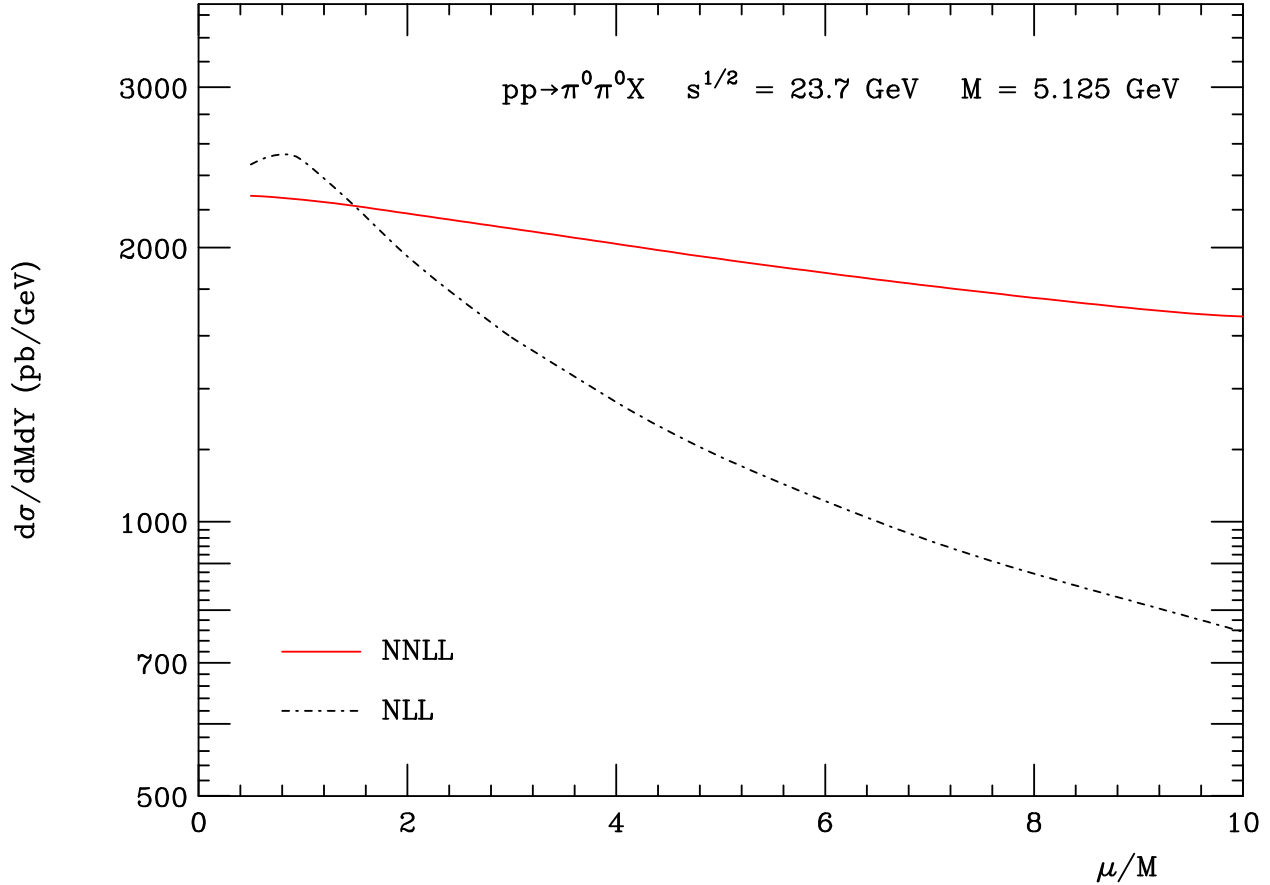,width=0.9\textwidth,angle=90}
\end{center}
\vspace*{-2cm}
\caption{\sf  Comparison of the scale dependence of the NLL resummed (dot-dashed) and the NNLL resummed (solid) cross 
sections for NA24 kinematics. We choose a pion pair invariant mass of $M=5.125$ GeV and show the variation of the 
cross sections as a function of $\mu/M$, where $\mu_R=\mu_F=\mu$. 
\label{fig:scale_dep}}
\vspace*{0.1cm}
\end{figure}

In order to assess the improvement in scale dependence in a more detailed way, we show in Fig.~\ref{fig:scale_dep}
results for the predicted cross section as a function of $\mu/M$ (where again $\mu_R=\mu_F\equiv\mu$), using a fixed 
pair invariant mass $M=5.125$~GeV, which corresponds to the left-most point in Fig.~\ref{fig:na24nnll}. 
The dot-dashed line corresponds to the variation of the NLL resummed cross section, where for 
$\xi_R$ in Eq.~(\ref{xi111}) we include only the first term in the exponent, i.e. $2b_0\as \ln(\mu_R^2/\hat m^2)$. 
This is the only term justified for a cross section resummed to this accuracy. We note that keeping this term in the
exponent or expanding the exponential to first order (as done in~\cite{Almeida:2009jt}) 
makes only a modest numerical difference. At NNLL, we include the full exponent 
$\xi_R$ in Eq.~(\ref{xi111}), keeping in mind the discussion following Eq.~(\ref{eq:scale1}). Our result for the scale 
variation of the NNLL resummed cross section is shown as a solid line in Fig.~\ref{fig:scale_dep}. One observes a 
very strong improvement when going from NLL to NNLL, with the NNLL resummed cross section rather flat
even out to scales as large as $\mu=10M$. 
%%%%%%%%%%%%%%%%%%%%%%%%%%%%%%%%%%%%%%%%%%%%%%%%%%%%%%%%
%%%%%%%%%%%%%%%%%%%%%%%%%%%%%%%%%%%%%%%%%%%%%%%%%%%%%%%%

\begin{figure}[t]
\begin{center}
\vspace*{-1.8cm}
\hspace*{-1.6cm}
\epsfig{figure=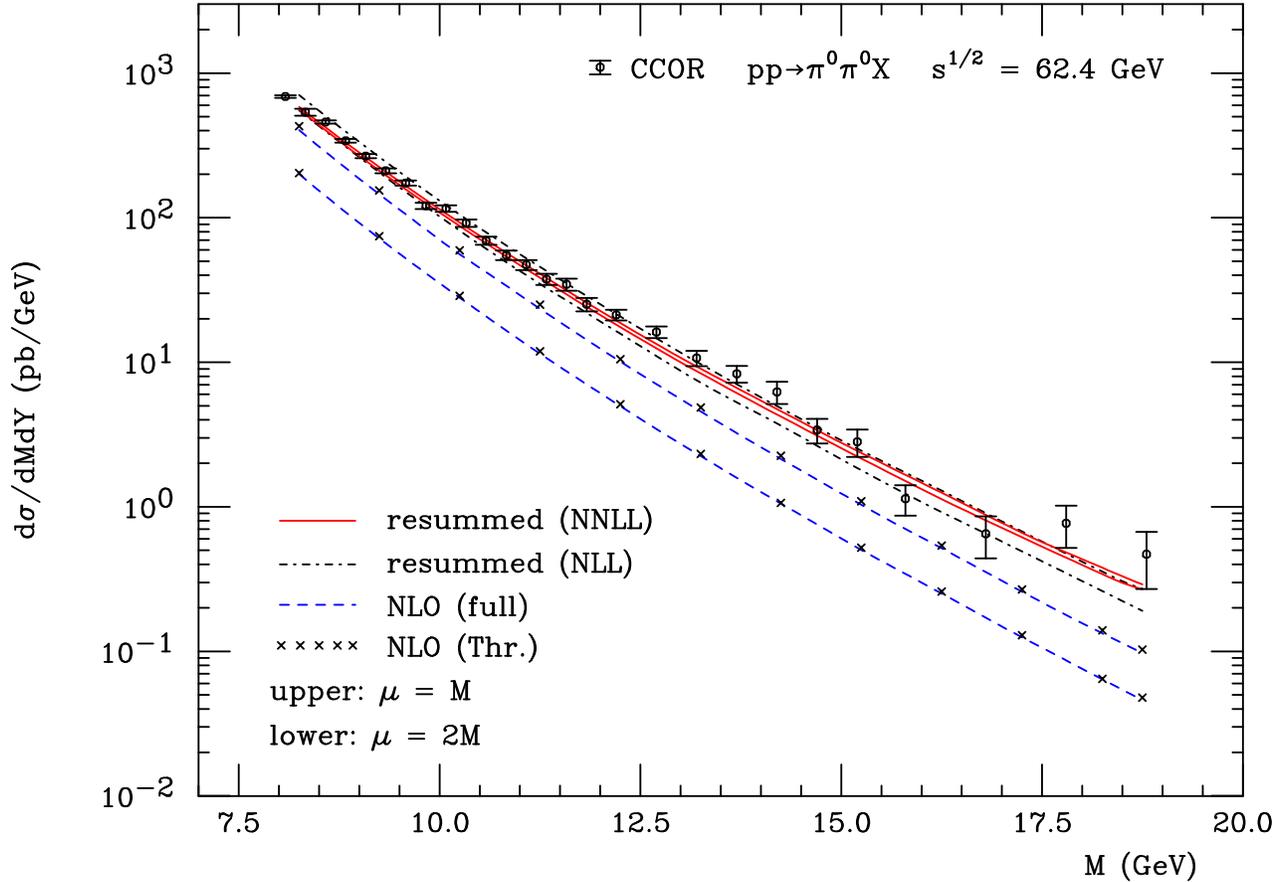,width=0.9\textwidth,angle=90}
\end{center}
\vspace*{-2cm}
\caption{\sf  Same as Fig.~\ref{fig:na24nnll}, but for $pp$ collisions at $\sqrt{S}=62.4$~GeV. 
The data are from CCOR~\cite{ccor}. 
\label{fig:ccor}}
\vspace*{0.1cm}
\end{figure}

Figure~\ref{fig:ccor} shows the comparison of our results to the CCOR data~\cite{ccor}. The main features of
the results are very similar to those in Fig.~\ref{fig:na24nnll}. Again the scale dependence is strongly 
reduced at NNLL. As a side remark we note that the new fragmentation functions of~\cite{DSS} also
help to achieve a much better description of the data than we found in our previous study~\cite{Almeida:2009jt}.
%%%%%%%%%%%%%%%%%%%%%%%%%%%%%%%%%%%%%%%%%%%%%%%%%%%%%%%%
%%%%%%%%%%%%%%%%%%%%%%%%%%%%%%%%%%%%%%%%%%%%%%%%%%%%%%%%

\begin{figure}[t]
\begin{center}
\vspace*{-1.8cm}
\hspace*{-1.6cm}
\epsfig{figure=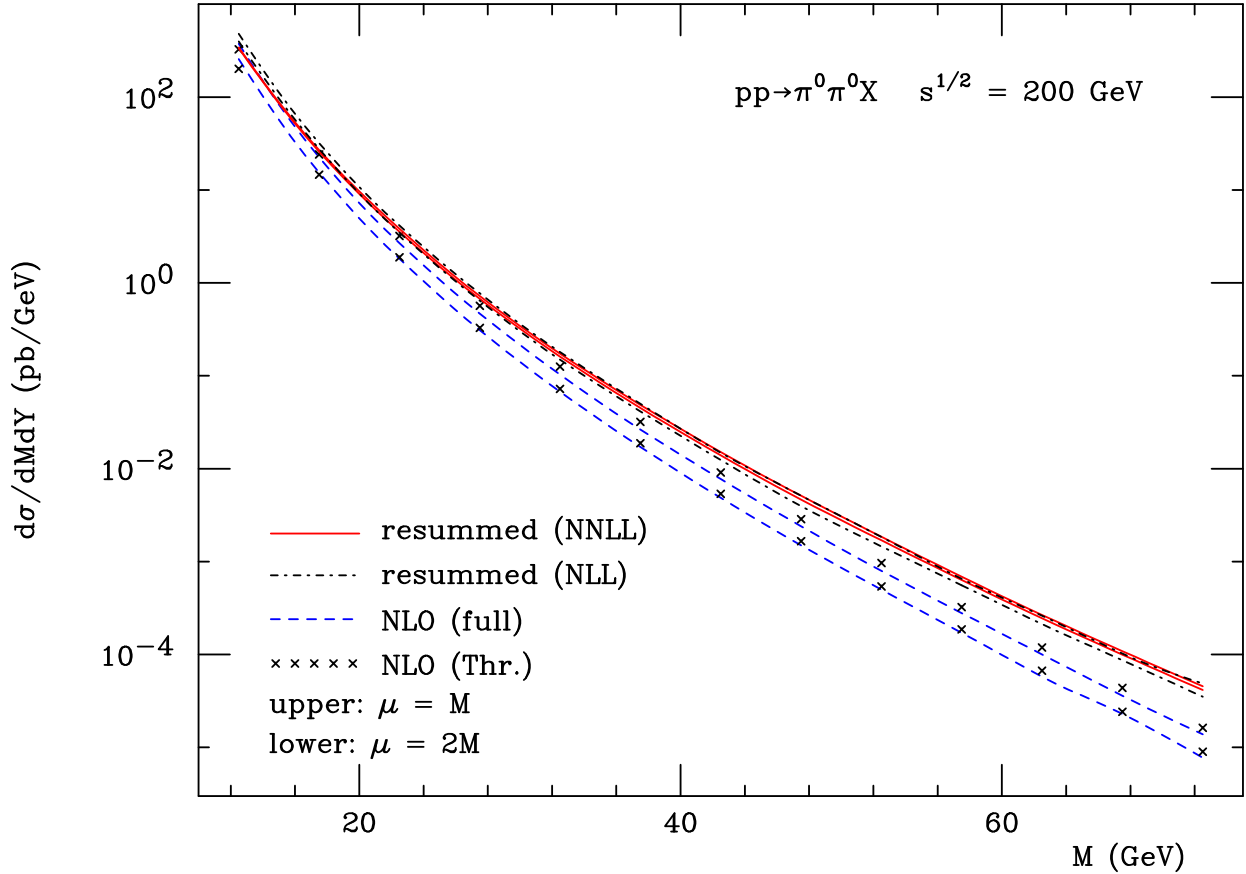,width=0.9\textwidth,angle=90}
\end{center}
\vspace*{-2cm}
\caption{\sf  Di-hadron production at RHIC at a center-of-mass energy of $\sqrt{S}=200$~GeV. The full 
NLO result (dashed) is shown in comparison to the NLO expansion of the resummed result (crosses). The solid 
line shows the NNLL resummed cross section. As before, we use the scales $\mu_R=\mu_F=M$ and $\mu_R=\mu_F=2M$. 
\label{fig:phenix}}
\vspace*{0.1cm}
\end{figure}

Finally, we consider di-hadron production in $pp$ collisions at RHIC with a c.m.s. energy of $\sqrt{S}=200$~GeV. 
For simplicity, we use the same cuts as for the NA24 experiment. In Fig.~\ref{fig:phenix}, we show our results for an 
invariant mass range of $M=10-75$~GeV. We find that at this energy the full NLO (dashed) and the NLO expansion 
of the resummed result (crosses) do not match quite as well as observed for fixed target scattering in Fig.~\ref{fig:na24nnll},
although the agreement is usually at the $10\%$ level or better. Threshold resummation again yields a sizable
enhancement over NLO, but the effects are somewhat smaller than in the fixed-target regime, since at RHIC's 
higher energy one is typically further away from threshold. Also here, the 
NNLL-resummed result is nearly within the NLL scale uncertainty band and shows a reduced scale dependence.
%%%%%%%%%%%%%%%%%%%%%%%%%%%%%%%%%%%%%%%%%%%%%%%%%%%%%%%%
%%%%%%%%%%%%%%%%%%%%%%%%%%%%%%%%%%%%%%%%%%%%%%%%%%%%%%%%

\section{Conclusions \label{sec5}}

We have extended the threshold resummation framework for di-hadron production in hadronic collisions,
$H_1 H_2\to h_1 h_2 X$, beyond the next-to-leading logarithmic level. To achieve this, we have determined
the first-order corrections to the hard-scattering function $H$ and the soft function $S$, which both are
matrices in color space. With these, it becomes possible to resum four towers of threshold logarithms in the 
perturbative series. In our numerical studies, we have found that the NNLL resummed results fall within the 
scale uncertainty band of the NLL resummed calculation. They also show a much reduced scale dependence. 

There are important further applications of our work. Of particular interest are di-jet, single-inclusive jet and 
single-inclusive hadron cross sections, all of which have much phenomenological relevance at present-day 
collider experiments. Given the promising results we have obtained for di-hadron production, we may 
expect that a similar resummation at NNLL for these reactions would also improve the theoretical QCD 
prediction. 

\section*{Acknowledgments} 

We are grateful to Leandro Almeida, Marco Stratmann, and Ilmo Sung for valuable discussions. 
This work was supported by the ``Bundesministerium f\"{u}r 
Bildung und Forschung'' (BMBF, grant no. 05P12VTCTG). The work of GS was supported in part by the National 
Science Foundation,  grants No. PHY-0969739 and No. PHY-1316617.

\begin{appendix}

\renewcommand{\theequation}{A.\arabic{equation}}
\setcounter{equation}{0} 

\section{The normalization of the soft function}

The soft matrix $S_{LI}$ in the resumed cross section in moment space, Eq.\ (\ref{resumm}), 
is computed as described in Ref.\ \cite{KOS}.   Its all-orders form is most conveniently exhibited in 
moment space, as the ratio of the moments of a fully eikonal cross section $\hat \sigma^{ab\rightarrow cd}_{LI}$
and four factorized jets, two to absorb the factorizing collinear singularities of the incoming parton lines, and two to 
absorb the collinear singularities of outgoing lines:
\be
\Big(S_{ab\to cd}\left( \as(\hat m/\bar{N}), \Delta \eta\right)\Big)_{LI}\ =\ 
\frac{\hat \sigma^{ab\rightarrow cd}_{LI} \left( \frac{\hat m}{N\mu_R}, \Delta \eta, 
\as(\mu_R),\varepsilon\right)}{\prod_{i=a,b} \tilde j_{\rm in}^{(i)}\left( \frac{\hat m}{N\mu_R},\as(\mu_R),\varepsilon \right)
\prod_{j=c,d} \tilde j_{\rm out}^{(j)}\left( \frac{\hat m}{N\mu_R},\as(\mu_R),\varepsilon \right)}\, .
\label{eq:soft-matrix}
\ee
As described in Refs.\ \cite{KS,KOS}, these ``in" and ``out" jets, $\tilde{j}_{\rm in}$ and $\tilde{j}_{\rm out}$, 
respectively are defined to match the collinear singularities and radiation phase space in the partonic threshold limit.

The explicit calculation of $(S_{ab\rightarrow cd})_{LI}$ at one loop as given here is equivalent to the procedure described in Sec.\ \ref{SF}.  
The functions on the right of (\ref{eq:soft-matrix}), as defined in detail below, are normalized and expanded according to
\bea
\hat \sigma^{ab\rightarrow cd}_{LI} \left( \frac{\hat m}{N\mu_R}, \Delta \eta, \as(\mu_R),\varepsilon \right)\ &=&\ 
(S_{ab\rightarrow cd}^{(0)})_{LI}\ +\ \frac{\alpha_s(\mu_R)}{\pi}\,
\hat \sigma^{ab\rightarrow cd \,(1)}_{LI}\ + \ {\cal O}(\alpha_s(\mu_R)^2)\, , \nn\\[2mm]
\tilde{j}_{\rm in}^{(i)} \left( \frac{\hat m}{N\mu_R},\as(\mu_R),\varepsilon \right) &=&\ 1\ +\ \frac{\alpha_s(\mu_R)}{\pi}\,
\tilde{j}_{\rm in}^{(i,1)}\ +\ {\cal O}(\alpha_s(\mu_R)^2)\, , \nn\\[2mm]
\tilde{j}_{\rm out}^{(j)} \left( \frac{\hat m}{N\mu_R},\as(\mu_R),\varepsilon \right) &=&\ 1\ +\ \frac{\alpha_s(\mu_R)}{\pi}\,
\tilde{j}_{\rm out}^{(j,1)}
\ +\ {\cal O}(\alpha_s(\mu_R)^2)\, ,
\label{eq:lo-SLI-and-jets}
\eea
where $S^{(0)}$ is the tree-level soft matrix, defined as in Eq.\ (\ref{eq:S0-trace}).    The first-order expansion of the soft matrix is thus,
\bea
\left(S_{ab\to cd}^{(1)}\right)_{LI}\ = \hat \sigma^{ab\rightarrow cd\, (1)}_{LI}\ - \ (S_{ab\rightarrow cd}^{(0)})_{LI} 
\left[ \,\sum_{i=a,b}\ \tilde{j}_{\rm in}^{(i,1)}\ +\ \sum_{j=c,d}\ \tilde{j}_{\rm out}^{(j,1)} \,\right ]\, .
\label{eq:eik-diff}
\eea
At any loop order, the collinear singularities of the eikonal cross section $\hat \sigma_{LI}$ match those of properly-defined incoming 
and outgoing jet functions.   At one loop, this will result in a finite soft function by simple cancellation in Eq.\ (\ref{eq:eik-diff}), as seen
in Sec.\ \ref{SF}.  That is, division by the regularized jet functions plays the role of the collinear factorization of the soft function.   
It also provides finite, factorizing corrections to the soft function, which depend on the definitions of the jets functions.    
Here we use jet functions defined directly from the eikonal resummations of Drell-Yan and double inclusive cross sections~\cite{Czakon:2009zw}.
The choices, defined below, match collinear singularities of the eikonal cross section, and have the advantage of being 
Lorentz and gauge invariant.   They differ from those made in Refs.\ \cite{KS,KOS} by finite terms, but the collinear structure 
is identical.   When restricted to the amplitude level, this is the same formalism that was implemented in Refs.\ \cite{twoloopad,Sterman:2002qn}.

To make the connection to the calculation of the soft function in this paper explicit, we recall that eikonal diagrams are generated by 
path-ordered exponentials with constant  velocities $\beta$, which we represent as
\beq
\Phi_{\beta}^{(f)}({\lambda}_2,{\lambda}_1;x)
=
P\exp\left(-ig\int_{{\lambda}_1}^{{\lambda}_2}d{\eta}\; {\beta}{\cdot} A^{(f)} ({\eta}{\beta}+x)\right)\, ,
\label{eq:wilson} 
\eeq
where superscript $f$ represents the color representation of the parton to which this ``Wilson line" corresponds.
In terms of these path-ordered exponentials, we define products corresponding to scattering, pair annihilation and 
pair creation.   For the case of $2\rightarrow 2$ scattering, the ends of two incoming and two outgoing Wilson 
lines are coupled locally by a constant color tensor ${\cal C}_I$,
\beqa
w_I^{(ab\rightarrow cd)}(x)_{\{j\}}
&=& 
\sum_{\{i\}}
\Phi_{\beta_d}^{(d)}(\infty,0;x)_{j_d,i_d}\; 
\Phi_{\beta_c}^{(c)}(\infty,0;x)_{j_c,i_c}\nonumber \\[2mm]
&\ &\times
\left( {\cal C}_I^{(ab\rightarrow cd)}\right)_{i_di_c,i_bi_a}\; 
\Phi_{\beta_a}^{(a)}(0,-\infty;x)_{i_a,j_a}
\Phi_{\beta_b}^{(b)}(0,-\infty;x)_{i_b,j_b}\, .
\label{eq:wivertex}
\eeqa 
For pair annihilation, two lines in conjugate representations that come from the infinite past are joined by 
a color singlet tensor, that is, a simple Kronecker delta,
\beqa
w_0^{(a\bar a)}(x)_{\{j\}}
&=& 
\sum_{\{i\}}
\left( \delta \right)_{i_a,i_{\bar{a}}}\; 
\Phi_{\beta_{\bar a}}^{(\bar a)}(0,-\infty;x)_{i_{\bar{a}},j_{\bar{a}}}
\Phi_{\beta_a}^{(a)}(0,-\infty;x)_{i_a,j_a}\, ,
\label{eq:w0vertex}
\eeqa
and similarly for pair creation, using color-conjugate lines that emerge from a point, and extend into the infinite future,
\beqa
\hat w_0^{(a\bar a)}(x)_{\{j\}}
&=& 
\sum_{\{i\}}
\Phi_{\beta_{\bar a}}^{(\bar a)}(\infty,0;x)_{i_{\bar{a}},j_{\bar{a}}}
\Phi_{\beta_a}^{(a)}(\infty,0;x)_{i_a,j_a}
\; \left( \delta \right)_{i_a,i_{\bar a}}\, .
\label{eq:w0vertex1}
\eeqa
In terms of these operators, the eikonal cross section is defined by
\bea
\hat \sigma^{ab\rightarrow cd}_{LI} \left( \frac{\hat m}{N\mu_R}, \Delta \eta, \as(\mu_R),\varepsilon
\right) \ &=& \ \int_0^1 d\tau\  \tau^{N-1}\ \int \frac{d y^0}{2\pi} \ e^{i\tau \hat m y^0}
\nn\\[1mm]
&\ & \times \ {\rm Tr}_{\{j\}} \;
\langle 0|\, \bar T\left(w_L^{(ab\rightarrow cd)}{}^\dagger\left ((y^0,\vec 0) \right)_{\{j\}} \right) \ T \left( w_I^{(ab\rightarrow cd)}(0)_{\{j\}} \right) |0\rangle 
\nn \\[2mm]
&=& \  \int_0^1 d\tau\  \tau^{N-1}\ \sum_\xi\ \delta(\tau \hat m- p_\xi^0)
\nn\\[1mm]
&\ & \times \ {\rm Tr}_{\{j\}} \;  \langle 0|\, \bar T\left( w_L^{(ab\rightarrow cd)}{}^\dagger 
\left(0\right)_{\{j\}} \right)  |\xi\rangle \, \langle \xi| T \left(  w_I^{(ab\rightarrow cd)}(0)_{\{j\}} \right) |0\rangle\, ,\nn\\
\eea
where $T$ represents time ordering, $\bar{T}$  anti-time ordering, and $p_\xi^0$ is the energy of  state $|\xi\rangle$. 
The in jet is defined {\it in terms of its square} in moment space as
\bea
\left( \tilde j_{\rm in}^{(a)}\left( \frac{\hat m}{N\mu_R},\as(\mu_R),\varepsilon \right) \right)^2\
&=& \   \int_0^1 d\tau\  \tau^{N-1}\ \sum_\xi\ \delta(\tau \hat m- p_\xi^0)
\nn \\[1mm]
&\ & \times \ {\rm Tr}_{\{j\}} \;  \langle 0|\, \bar T\left( w_0^{(a\bar a)}{}^\dagger 
\left(0\right)_{\{j\}} \right)  |\xi\rangle \, \langle \xi| T \left(  w_0^{(a\bar a)}(0)_{\{j\}} \right) |0\rangle\, .
\nn \\
\eea
With this choice, $\left(\tilde{j}^{(a)}_{\rm in}\right)^2$ is exactly the eikonal Drell-Yan cross section. It was computed to two 
loops in Ref.\ \cite{Belitsky:1998tc}.  The out jet is defined by the same integrals but with the pair of incoming Wilson 
lines of the operator $w_0(x)$ replaced by the outgoing pair in $\hat w_0(x)$, corresponding to double 
inclusive annihilation \cite{Sterman:2006hu}:
\bea
\left( \tilde j_{\rm out}^{(c)}\left( \frac{\hat m}{N\mu_R},\as(\mu_R),\varepsilon \right) \right )^2\
&=& \  \int_0^1 d\tau\  \tau^{N-1}\ \sum_\xi\ \delta(\tau \hat m- p_\xi^0)
\nn\\[1mm]
&\ & \times \  {\rm Tr}_{\{j\}} \;  \langle 0|\, \bar T\left(\hat w_0^{(c\bar c)}{}^\dagger \left(0\right)_{\{j\}} \right)  
|\xi\rangle \, \langle \xi| T \left( \hat w_0^{(c\bar c)}(0)_{\{j\}} \right) |0\rangle\, .
\nn\\
\eea
It is easy to confirm explicitly in Ref.\ \cite{Belitsky:1998tc} that the calculation of this quantity depends only on the inner 
products $\beta_a\cdot \beta_{\bar a}$ so that the full two-loop calculation and renormalization of this operator is the same 
for outgoing as for incoming eikonal jets.  

The resummation of logarithms of $N$ in this cross section leads precisely to the functions $\ln \Delta_i^N$ in Eq.\ (\ref{DfctNLL}), 
which summarize factoring NNLL dependence on the moment variable $N$, as confirmed recently 
in Ref.\ \cite{Sterman:2013nya}.   We note, however, that in the NNLL exponentiation as implemented into the expression for the 
functions $\Delta_i^N$ in Eq.\ (\ref{Dfct}), the Drell-Yan soft function is treated as an overall prefactor evaluated at the hard 
scale $\hat m$, rather than at $\hat m/N$.    Logarithms at NNLL that are associated with this shift are already incorporated into 
the exponent by use of the relation \cite{Sterman:2013nya}
\bea
S(\as(\hat m/N))\ =\  S(\as(\hat m))\exp \left[ -\int_{\hat m/N}^{\hat m} \frac{d\mu}{{\mu}} \frac{\partial \ln S(\as(\mu))}{\partial \mu} \right]\, .
\eea
To match logarithms associated with these factors consistently we include in our definition of $\Delta_i^N$ in Eq.\ (\ref{Dfct})
an extra factor of $R_i=1-(3\as/4\pi)A_i^{(1)}\zeta(2)$, Eq.~(\ref{Ri}), to account for our definitions of the in- and out-jet functions in terms 
of Drell-Yan and double inclusive cross sections. The combined factors for all four jet functions match the $\pi^2$ contribution 
in~(\ref{R3}), which in turn arises from the explicit $\pi^2$ terms in the integrals $dI_{ij}/d\hat\tau$ in~(\ref{eq:ls}). 

\end{appendix}

%%%%%%%%%%%%%%%%%%%%%%%%%%%%%%%%%%%%%%%%%%%%%%%%%%%%%%%%
%%%%%%%%%%%%%%%%%%%%%%%%%%%%%%%%%    References    %%%%%%%%%%%%%%%
%%%%%%%%%%%%%%%%%%%%%%%%%%%%%%%%%%%%%%%%%%%%%%%%%%%%%%%%

%\vspace*{-3mm}


\newpage
\begin{thebibliography}{99}

%\vspace*{-3mm}
\bibitem{dyresum} G.~F.~Sterman, Nucl.\ Phys.\ B {\bf 281}, 310 (1987);
S.~Catani and L.~Trentadue, Nucl.\ Phys.\ B {\bf 327}, 323 (1989);
Nucl.\ Phys.\ B {\bf 353}, 183 (1991).

\bibitem{KS} 
N.~Kidonakis and G.~F.~Sterman,
%``Resummation for QCD hard scattering,''
Nucl.\ Phys.\ B {\bf 505}, 321 (1997)
[arXiv:hep-ph/9705234].

\bibitem{BCMN}  R.~Bonciani, S.~Catani, M.~L.~Mangano and P.~Nason,
  %``Sudakov resummation of multiparton QCD cross sections,''
  Phys.\ Lett.\  B {\bf 575}, 268 (2003)
  [arXiv:hep-ph/0307035].
  %%CITATION = PHLTA,B575,268;%%
  
\bibitem{Catani:2003zt} 
  S.~Catani, D.~de Florian, M.~Grazzini and P.~Nason,
  %``Soft gluon resummation for Higgs boson production at hadron colliders,''
  JHEP {\bf 0307}, 028 (2003)
  [hep-ph/0306211].
  
\bibitem{Ahrens:2008nc} 
  V.~Ahrens, T.~Becher, M.~Neubert and L.~L.~Yang,
  %``Renormalization-Group Improved Prediction for Higgs Production at Hadron Colliders,''
  Eur.\ Phys.\ J.\ C {\bf 62}, 333 (2009)
  [arXiv:0809.4283 [hep-ph]];  
%   V.~Ahrens, T.~Becher, M.~Neubert and L.~L.~Yang,
  %``Updated Predictions for Higgs Production at the Tevatron and the LHC,''
  Phys.\ Lett.\ B {\bf 698}, 271 (2011)
  [arXiv:1008.3162 [hep-ph]].

\bibitem{Bonvini:2014joa} 
  M.~Bonvini and S.~Marzani,
  %``Resummed Higgs cross section at N$^{3}$LL,''
  JHEP {\bf 1409}, 007 (2014)
  [arXiv:1405.3654 [hep-ph]].

\bibitem{Catani:2014uta} 
  S.~Catani, L.~Cieri, D.~de Florian, G.~Ferrera and M.~Grazzini,
  %``Threshold resummation at N$^3$LL accuracy and soft-virtual cross sections at N$^3$LO,''
  Nucl.\ Phys.\ B {\bf 888}, 75 (2014)
  [arXiv:1405.4827 [hep-ph]].
  
\bibitem{Anastasiou:2014vaa} 
  C.~Anastasiou, C.~Duhr, F.~Dulat, E.~Furlan, T.~Gehrmann, F.~Herzog and B.~Mistlberger,
  %``Higgs boson gluonÐfusion production at threshold in $N^3LO$ $QCD$,''
  Phys.\ Lett.\ B {\bf 737}, 325 (2014)
  [arXiv:1403.4616 [hep-ph]].  

\bibitem{KOS} N.~Kidonakis, G.~Oderda and G.~F.~Sterman,
Nucl.\ Phys.\ B {\bf 525}, 299 (1998) [arXiv:hep-ph/9801268];
Nucl.\ Phys.\ B {\bf 531}, 365 (1998) [arXiv:hep-ph/9803241].

\bibitem{KO1}  
  N.~Kidonakis and J.~F.~Owens,
  %``Effects of higher-order threshold corrections in high-E(T) jet
  %production,''
  Phys.\ Rev.\  D {\bf 63}, 054019 (2001)
  [arXiv:hep-ph/0007268].
  %%CITATION = PHRVA,D63,054019;%%

\bibitem{Catani:2013vaa} 
  S.~Catani, M.~Grazzini and A.~Torre,
  %``Soft-gluon resummation for single-particle inclusive hadroproduction at high transverse momentum,''
  Nucl.\ Phys.\ B {\bf 874}, 720 (2013)
  [arXiv:1305.3870 [hep-ph]].  
  
\bibitem{Kidonakis:2003qe}
   N.~Kidonakis and R.~Vogt,
  %``Next-to-next-to-leading order soft gluon corrections in top quark hadroproduction,''
  Phys.\ Rev.\ D {\bf 68}, 114014 (2003)
  [hep-ph/0308222]; Phys.\ Rev.\ D {\bf 78}, 074005 (2008)
  [arXiv:0805.3844 [hep-ph]];
N.~Kidonakis,
  %``Next-to-next-to-leading soft-gluon corrections for the top quark cross section and transverse momentum distribution,''
  Phys.\ Rev.\ D {\bf 82}, 114030 (2010)
  [arXiv:1009.4935 [hep-ph]];
%N.~Kidonakis,
  %``NNNLO soft-gluon corrections for the top-antitop pair production cross section,''
  Phys.\ Rev.\ D {\bf 90}, 014006 (2014)
  [arXiv:1405.7046 [hep-ph]];
%N.~Kidonakis,
  %``NNNLO soft-gluon corrections for the top-quark $p_T$ and rapidity distributions,''
  arXiv:1411.2633 [hep-ph].  
  
\bibitem{Beneke:2011mq}
 M.~Beneke, P.~Falgari and C.~Schwinn,
  %``Soft radiation in heavy-particle pair production: All-order colour structure and two-loop anomalous dimension,''
  Nucl.\ Phys.\ B {\bf 828}, 69 (2010)
  [arXiv:0907.1443 [hep-ph]];
M.~Beneke, P.~Falgari, S.~Klein and C.~Schwinn,
  %``Hadronic top-quark pair production with NNLL threshold resummation,''
  Nucl.\ Phys.\ B {\bf 855}, 695 (2012)
  [arXiv:1109.1536 [hep-ph]];
  M.~Beneke, Y.~Kiyo and K.~Schuller,
  %``Third-order correction to top-quark pair production near threshold I. Effective theory set-up and matching coefficients,''
  arXiv:1312.4791 [hep-ph].
  
\bibitem{Czakon:2009zw} 
  M.~Czakon, A.~Mitov and G.~F.~Sterman,
  %``Threshold Resummation for Top-Pair Hadroproduction to Next-to-Next-to-Leading Log,''
  Phys.\ Rev.\ D {\bf 80}, 074017 (2009)  [arXiv:0907.1790 [hep-ph]].

\bibitem{Yang:2014hya} 
  V.~Ahrens, A.~Ferroglia, M.~Neubert, B.~D.~Pecjak and L.~L.~Yang,
  %``Renormalization-Group Improved Predictions for Top-Quark Pair Production at Hadron Colliders,''
  JHEP {\bf 1009}, 097 (2010)
  [arXiv:1003.5827 [hep-ph]];
  %``RG-improved single-particle inclusive cross sections and forward-backward asymmetry in $t\bar t$ production at hadron colliders,''
  JHEP {\bf 1109}, 070 (2011)
  [arXiv:1103.0550 [hep-ph]];
  L.~L.~Yang, C.~S.~Li, J.~Gao and J.~Wang,
  %``NNLL momentum-space threshold resummation in direct top quark production at the LHC,''
  arXiv:1409.6959 [hep-ph].
  
\bibitem{Kidonakis:2010tc} 
  N.~Kidonakis,
  %``NNLL resummation for s-channel single top quark production,''
  Phys.\ Rev.\ D {\bf 81}, 054028 (2010)
  [arXiv:1001.5034 [hep-ph]];
  Phys.\ Rev.\ D {\bf 82}, 054018 (2010)
  [arXiv:1005.4451 [hep-ph]];
Phys.\ Rev.\ D {\bf 83}, 091503 (2011)
  [arXiv:1103.2792 [hep-ph]].
  
\bibitem{Beneke:2010da} 
  M.~Beneke, P.~Falgari and C.~Schwinn,
  %``Threshold resummation for pair production of coloured heavy (s)particles at hadron colliders,''
  Nucl.\ Phys.\ B {\bf 842}, 414 (2011)
  [arXiv:1007.5414 [hep-ph]].  
  
\bibitem{Beenakker:2013mva} 
  W.~Beenakker {\it et al.},
  %``Towards NNLL resummation: hard matching coefficients for squark and gluino hadroproduction,''
  JHEP {\bf 1310}, 120 (2013)
  [arXiv:1304.6354 [hep-ph]];
%\bibitem{Beenakker:2014sma} 
  W.~Beenakker {\it et al.},
%  , C.~Borschensky, M.~Kr\"{a}mer, A.~Kulesza, E.~Laenen, V.~Theeuwes and S.~Thewes,
  %``NNLL resummation for squark and gluino production at the LHC,''
  arXiv:1404.3134 [hep-ph].   
  
\bibitem{Broggio:2014hoa} 
  A.~Broggio, A.~Ferroglia, B.~D.~Pecjak and Z.~Zhang,
  %``NNLO hard functions in massless QCD,''
  JHEP {\bf 1412}, 005 (2014)  [arXiv:1409.5294 [hep-ph]].

\bibitem{Almeida:2009jt} L.~G.~Almeida, G.~F.~Sterman and W.~Vogelsang,
  %``Threshold Resummation for Di-hadron Production in Hadronic Collisions,''
  Phys.\ Rev.\ D {\bf 80}, 074016 (2009)
  [arXiv:0907.1234 [hep-ph]].
  
\bibitem{na24} C.~De Marzo {\it et al.} [NA24 Collaboration],  
%``Measurement of the production of high mass gamma gamma, pi0 pi0, 
% and gamma 
%pi0 pairs in pi- p, pi+ p, and p p collisions at 300-GeV/c,''
Phys.\ Rev.\  D {\bf 42}, 748 (1990).

\bibitem{e711}
  H.~B.~White {\it et al.} [E711 Collaboration],
  %``Massive hadron pair production by 800-geV/c protons on nuclear targets,''
  Phys.\ Rev.\  D {\bf 48}, 3996 (1993);
  %%CITATION = PHRVA,D48,3996;%%
H.~B.~White, {\it A Study of angular dependence in parton-parton 
scattering from massive hadron pair production}, PhD Thesis Florida State U.,
FERMILAB-THESIS-1991-39, FSU-HEP-910722, UMI-92-02321, 1991. 

\bibitem{e706}
M.~Begel [E706 Collaboration],
{\it Production of high mass pairs of direct photons and neutral mesons in a
Tevatron fixed target experiment}, PhD Thesis Univ. of Rochester, FERMILAB-THESIS-1999-05, 
UMI-99-60725, 1999.

\bibitem{ccor}
  A.~L.~S.~Angelis {\it et al.}  [CCOR Collaboration],
  %``RESULTS ON CORRELATIONS AND JETS IN HIGH TRANSVERSE MOMENTUM $p p$
  %COLLISIONS AT THE CERN ISR,''
Nucl.\ Phys.\ B {\bf 209}, 284 (1982). 

\bibitem{Kunszt:1993sd} 
  Z.~Kunszt, A.~Signer and Z.~Trocsanyi,
  %``One loop helicity amplitudes for all 2 ---> 2 processes in QCD and N=1 supersymmetric Yang-Mills theory,''
  Nucl.\ Phys.\ B {\bf 411}, 397 (1994)
  [hep-ph/9305239].
  
\bibitem{Bern:1990cu} Z.~Bern and D.~A.~Kosower,
%Efficient calculation of one loop QCD amplitudes
Phys.\ Rev.\ Lett.\ {\bf 66}, 1669 (1991).

\bibitem{Bern:1991aq} Z.~Bern and D.~A.~Kosower,
%The Computation of loop amplitudes in gauge theories
Nucl.\ Phys.\ B {\bf 379}, 451 (1992).
  
\bibitem{Kelley:2010fn} 
  R.~Kelley and M.~D.~Schwartz,
  %``1-loop matching and NNLL resummation for all partonic 2 to 2 processes in QCD,''
  Phys.\ Rev.\ D {\bf 83}, 045022 (2011)
  [arXiv:1008.2759 [hep-ph]].

\bibitem{Cacciari:2001cw} 
M.~Cacciari and S.~Catani, 
%``Soft-Gluon Resummation for the Fragmentation of Light and Heavy Quarks at
%Large x,''  
Nucl.\ Phys.\  B {\bf 617}, 253 (2001) [arXiv:hep-ph/0107138].

%\cite{Sterman:2006hu}
\bibitem{Sterman:2006hu}
  G.~F.~Sterman and W.~Vogelsang,
  %``Crossed threshold resummation,''
  Phys.\ Rev.\  D {\bf 74}, 114002 (2006)
  [arXiv:hep-ph/0606211].
  %%CITATION = PHRVA,D74,114002;%%
  
\bibitem{Vogt:2000ci} A.~Vogt,
  %``Next-to-next-to-leading logarithmic threshold resummation for deep inelastic scattering and the Drell-Yan process,''
  Phys.\ Lett.\ B {\bf 497}, 228 (2001)
  [hep-ph/0010146];
S.~Moch, J.~A.~M.~Vermaseren and A.~Vogt,
  %``Higher-order corrections in threshold resummation,''
  Nucl.\ Phys.\ B {\bf 726}, 317 (2005)  [hep-ph/0506288].

\bibitem{KT} J.~Kodaira and L.~Trentadue, 
%%``Summing Soft Emission In QCD,''
Phys.\ Lett.\ B {\bf 112}, 66 (1982); Phys.\ Lett.\ B {\bf 123}, 
335 (1983);\\ S.~Catani, E.~D'Emilio and L.~Trentadue,
%%``THE GLUON FORM-FACTOR TO HIGHER ORDERS: GLUON GLUON ANNIHILATION AT SMALL
  %Q-TRANSVERSE,'
Phys.\ Lett.\ B {\bf 211}, 335 (1988).

\bibitem{Moch:2004pa} S.~Moch, J.~A.~M.~Vermaseren and A.~Vogt,
  %``The Three loop splitting functions in QCD: The Nonsinglet case,''
  Nucl.\ Phys.\ B {\bf 688}, 101 (2004)
  [hep-ph/0403192].

\bibitem{Catani:2001ic} S.~Catani, D.~de Florian and M.~Grazzini,
%``Higgs production in hadron collisions: Soft and virtual QCD corrections  
%at NNLO,'' 
JHEP {\bf 0105}, 025 (2001) [arXiv:hep-ph/0102227].
 
\bibitem{Harlander:2001is} R.~V.~Harlander and W.~B.~Kilgore,
%``Soft and virtual corrections to p p --> H + X at NNLO,''  
Phys.\ Rev.\  D {\bf 64}, 013015 (2001)  [arXiv:hep-ph/0102241].

\bibitem{eric} T.~O.~Eynck, E.~Laenen and L.~Magnea,  
%``Exponentiation of the Drell-Yan cross section near partonic threshold  in
%the DIS and MS-bar schemes,''  
JHEP {\bf 0306}, 057 (2003) [arXiv:hep-ph/0305179];
E.~Laenen and L.~Magnea,  
%``Threshold resummation for electroweak annihilation from DIS data,'' 
Phys.\ Lett.\  B {\bf 632}, 270 (2006) [arXiv:hep-ph/0508284].

\bibitem{Tarasov:1980au} O.~V.~Tarasov, A.~A.~Vladimirov and A.~Yu.~Zharkov,
%The Gell-Mann-Low Function of QCD in the Three Loop Approximation
Phys.\ Lett.\ B {\bf 93}, 429 (1980).

\bibitem{Larin:1993tp} S.~A.~Larin and J.~A.~M.~Vermaseren,
  %``The Three loop QCD Beta function and anomalous dimensions,''
  Phys.\ Lett.\ B {\bf 303}, 334 (1993)
  [hep-ph/9302208].

\bibitem{msj} M.~Sjodahl,
  %``Color structure for soft gluon resummation - a general recipe,''
  JHEP {\bf 0909}, 087 (2009)
  [arXiv:0906.1121 [hep-ph]].

\bibitem{twoloopad}
S.~M.~Aybat, L.~J.~Dixon and G.~F.~Sterman,
Phys.\ Rev.\ Lett.\  {\bf 97}, 072001 (2006)  [arXiv:hep-ph/0606254];  
Phys.\ Rev.\  D {\bf 74}, 074004 (2006)  [arXiv:hep-ph/0607309];  
{\it see also:} E.~Gardi and L.~Magnea,  
JHEP {\bf 0903}, 079 (2009) [arXiv:0901.1091 [hep-ph]];
T.~Becher and M.~Neubert,  
%``On the Structure of Infrared Singularities of Gauge-Theory Amplitudes,''
JHEP {\bf 0906}, 081 (2009) [arXiv:0903.1126 [hep-ph]].  
%%CITATION = JHEPA,0906,081;%%

\bibitem{ddfwv}
  D.~de Florian and W.~Vogelsang,
  %``Threshold resummation for the inclusive hadron cross-section in p p
  %collisions,''
  Phys.\ Rev.\  D {\bf 71}, 114004 (2005)
  [arXiv:hep-ph/0501258]; 

\bibitem{deFlorian:2007fv} D.~de Florian and W.~Vogelsang,
  %``Resummed cross-section for jet production at hadron colliders,''
  Phys.\ Rev.\ D {\bf 76}, 074031 (2007)
  [arXiv:0704.1677 [hep-ph]];
 D.~de Florian, P.~Hinderer, A.~Mukherjee, F.~Ringer and W.~Vogelsang,
  %``Approximate next-to-next-to-leading order corrections to hadronic jet production,''
  Phys.\ Rev.\ Lett.\  {\bf 112}, 082001 (2014)
  [arXiv:1310.7192 [hep-ph]].
  
\bibitem{Laenen:2000ij} 
  E.~Laenen, G.~F.~Sterman and W.~Vogelsang,
  %``Recoil and threshold corrections in short distance cross-sections,''
  Phys.\ Rev.\ D {\bf 63}, 114018 (2001)
  [hep-ph/0010080].

\bibitem{KSV} A.~Kulesza, G.~F.~Sterman and W.~Vogelsang,
  %``Joint resummation in electroweak boson production,''
  Phys.\ Rev.\  D {\bf 66}, 014011 (2002)
  [arXiv:hep-ph/0202251].

\bibitem{Glover:2002gz} N.~E.~W.~Glover,
  %``Progress in NNLO calculations for scattering processes,''
  Nucl.\ Phys.\ Proc.\ Suppl.\  {\bf 116}, 3 (2003)
  [hep-ph/0211412].

\bibitem{Sterman:2002qn} G.~F.~Sterman and M.~E.~Tejeda-Yeomans,
  %``Multiloop amplitudes and resummation,''
  Phys.\ Lett.\ B {\bf 552}, 48 (2003)
  [hep-ph/0210130].
  
\bibitem{Catani:1998bh} 
  S.~Catani,
  %``The Singular behavior of QCD amplitudes at two loop order,''
  Phys.\ Lett.\ B {\bf 427}, 161 (1998)
  [hep-ph/9802439].  

\bibitem{WvN} W.~L.~van Neerven,  
%``Dimensional Regularization Of Mass And Infrared Singularities In Two Loop 
%On-Shell Vertex Functions,'' 
Nucl.\ Phys.\  B {\bf 268}, 453 (1986).

\bibitem{Catani:1996yz} S.~Catani, M.~L.~Mangano, P.~Nason and L.~Trentadue,
  %``The Resummation of Soft Gluon in Hadronic Collisions,''
  Nucl.\ Phys.\  B {\bf 478}, 273 (1996)
  [arXiv:hep-ph/9604351].
  %%CITATION = NUPHA,B478,273;%%

\bibitem{Owens:2001rr}
  J.~F.~Owens,
  %``A next-to-leading-order study of dihadron production,''
  Phys.\ Rev.\  D {\bf 65}, 034011 (2002)
  [arXiv:hep-ph/0110036].

\bibitem{cteq6} W.~K.~Tung, H.~L.~Lai, A.~Belyaev, J.~Pumplin, 
D.~Stump and C.~P.~Yuan, 
%``Heavy quark mass effects in deep inelastic scattering and global QCD
%analysis,''  
JHEP {\bf 0702}, 053 (2007) [arXiv:hep-ph/0611254].

\bibitem{DSS} D.~de Florian, R.~Sassot, M.~Epele, R.~J.~Hernandez-Pinto and M.~Stratmann,
  %``Parton-to-Pion Fragmentation Reloaded,''
  arXiv:1410.6027 [hep-ph], Phys. Rev. D (to appear).
  
%\cite{Belitsky:1998tc}
\bibitem{Belitsky:1998tc} 
  A.~V.~Belitsky,
  %``Two loop renormalization of Wilson loop for Drell-Yan production,''
  Phys.\ Lett.\ B {\bf 442}, 307 (1998)
  [hep-ph/9808389].
  %%CITATION = HEP-PH/9808389;%%
  
  %\cite{Sterman:2013nya}
\bibitem{Sterman:2013nya} 
  G.~F.~Sterman and M.~Zeng,
  %``Quantifying Comparisons of Threshold Resummations,''
  JHEP {\bf 1405}, 132 (2014)
  [arXiv:1312.5397 [hep-ph]].
        
\end{thebibliography}
\end{document}